\documentclass{aa}
\usepackage{natbib,epsfig,txfonts,graphicx,color,amssymb}
\definecolor{Blue}{rgb}{0.06,0.20,0.93}
\bibpunct{(}{)}{;}{a}{}{,}

\def\SiIIIa {Si III $\lambda$4553\,}

\def\SiIIIa{Si~III~$\lambda$4552\,}

\def\OIII{O~III $\lambda$5591\,}

\def\Mdot {$\dot M$\,}

\def\Yhe {$Y_{\rm He}$\,} 
\def\kms {km~s$^{\rm -1}$\,}

\def\Minit {$M^{init}_{evol}$\,}
\def\Teff {$T_{\rm eff}$\,}

\def\vinit {$v_{init}$\,}
\def\vsini {$v \sin i$\,}
\def\vcrit {$v_{crit}$\,} 
\def\logg {$\log g$\,}

\def\vmic {$v_{\rm mic}$\,}

\def\eb   {$\Theta_{RT}$\,}

\def\Msun {$M_\odot$\,}

\def \Teffe {T_{\rm eff}}

\def \ao{\frac{1}{\alpha'}}

\begin{document}
\title{Spectroscopic and physical parameters of Galactic O-type stars.}
\subtitle{II. Observational constraints on projected rotational and extra 
broadening velocities as a function of fundamental parameters and 
stellar evolution.
\thanks{Based on observations collected at the European Organisation for
Astronomical Research in the Southern Hemisphere, Chile, under programme 
ID 072.D-0196}}

\author{N. Markova\inst{1}, J. Puls\inst{2}, S. Sim\'on-D\'iaz\inst{3,4},
 A. Herrero\inst{3,4}, H. Markov\inst{1}, N. Langer\inst{5} }
\offprints{N. Markova,\\ \email{nmarkova@astro.bas.bg}}

\institute{Institute of Astronomy with  NAO, BAS, 
	P.O. Box 136, 4700 Smolyan, Bulgaria\\ 
	\email{nmarkova@astro.bas.bg}
\and  Universit\"{a}ts-Sternwarte,
	Scheinerstrasse 1, D-81679 M\"unchen, Germany\\
	\email{uh101aw@usm.uni-muenchen.de}
\and Instituto de Astrof\'isica de Canarias, 
      E38200 La Laguna, Tenerife, Spain.\\
\email{ssimon@iac.es}, 
\email{ahd@iac.es}
\and Departamento de Astrof\'isica, 
     Universidad de La Laguna, E-38205 La Laguna, Tenerife, Spain\\
     \email{ahd@iac.es}
\and  Argelander-Institut f\"ur Astronomie, Bonn University, D-53121 Bonn, Germany \\
\email{nlanger@astro.uni-bonn.de}.}
\date{Received; Accepted }
\abstract
%
{Rotation is of key importance for the evolution of massive star, including 
their fate as supernovae or Gamma-ray bursts. However, the rotational velocities  of OB 
stars are difficult to determine.}
{Based on our own data for 31 Galactic O stars and incorporating 
similar data for 86 OB supergiants from the literature, we aim at
investigating the properties of rotational and extra line-broadening 
as a function of stellar parameters and at testing model predictions about 
the evolution of stellar rotation.
}
{Fundamental stellar parameters were  determined by means of the code
FASTWIND. Projected rotational and extra broadening velocities, \vsini\, 
and \eb, originate from  a combined Fourier transform + goodness-of-fit 
method. Model calculations published previously were used to estimate 
the initial evolutionary masses, \Minit.
}
{The sample O stars  with \Minit $\gtrsim$ 50~\Msun\, rotate with 
less that  26\% of their break-up velocity, and they also lack slow rotators 
(\vsini~$\lesssim$~50~\kms). For the more massive stars (\Minit$\ge$35~\Msun) 
on the hotter side of the bi-stability jump, the observed and predicted 
rotational rates agree quite well; for those on the cooler 
side of the jump, the measured velocties are systematically higher  than the 
predicted ones. In general, the derived \eb\, values decrease toward cooler 
\Teff, whilst for  later evolutionary phases they appear, at the same \vsini, 
higher for high-mass stars than for low-mass ones. None of the sample stars 
shows \eb $\ge $~110~\kms. For the majority of the more massive stars, extra 
broadening either  dominates or is in strong competition with rotation.
}
{For OB stars of solar metallicity, extra broadening is important and
has to be accounted for in the analysis. When appearing at or close to
the zero-age main sequence, most of the single and more massive stars
rotate slower than previously thought. 
Model predictions for the
evolution of rotation in hot massive stars may need to be updated.
}

\titlerunning{Rotational and extra broadening}
\authorrunning{N. Markova et al.}

\maketitle

%
\section{Introduction}

Rotation has been initially considered as a secondary effect for the evolution 
of massive stars. This view was challenged later on by a number of significant 
discrepancies between model predictions and observations (see, e.g., 
\citealt{maeder95} and references therein)  which,  on the contrary, suggest 
that rotation may have an impact comparable to that of mass loss, a possibility 
that has been entirely confirmed by the new generation of evolutionary models 
that account for stellar rotation  \citep{mm00,brott}.

Although of critical importance for   constraining evolutionary predictions, the 
rotational rates of massive OB stars are not clearly determined yet, and  one of the 
major difficulties is related to the additional  broadening in the spectrum of 
these stars \citep{slettebak, rozhal, CE77, penny96, howarth97},which is typically 
designated as macroturbulence.

Even though the nature and the origin of macroturbulence is still unclear, its 
presence in the atmospheres of hot massive stars may have important implications 
for our knowledge of the physics of these objects. In particular, through its 
effect on the projected rotational rates, macroturbulence may affect the 
interpretation of stellar evolution calculations, providing  a clue for understanding 
the lack of consistency between model predictions and observations regarding  
the role of rotational mixing in enriching surface N abundances in massive OB 
stars \citep{hunter08, hunter09, gonzalez12}. Through its effect on the shape 
and width of spectral lines, it might also modify stellar properties derived 
by means of model atmosphere analyses.
\begin{table*}
\begin{center}
\caption{Derived stellar properties of our basic O-star sample} 
\label{sample}
\tabcolsep1.8mm
\begin{tabular}{llcccccc}
\hline
\hline
\multicolumn{1}{l}{Star}
&\multicolumn{1}{l}{SpT}
&\multicolumn{1}{c}{\vsini\, (PF)}
&\multicolumn{1}{l}{\vsini\, (FT+GOF)}
&\multicolumn{1}{c}{\eb }
&\multicolumn{1}{c}{\Teff}
&\multicolumn{1}{c}{\logg}
&\multicolumn{1}{c}{\Minit}\\
\multicolumn{1}{l}{}
&\multicolumn{1}{l}{}
&\multicolumn{1}{c}{ [\kms]}
&\multicolumn{1}{c}{ [\kms]}
&\multicolumn{1}{c}{ [\kms]}
&\multicolumn{1}{c}{ [kK]}
&\multicolumn{1}{c}{}
&\multicolumn{1}{c}{ [\Msun]}
\\
\hline
\hline
\\
HD~64568        &O3 V((f*))    &90    &55   &96   &46.5   &3.90  &80 \\
HD~46223        &O4~V((f))     &100   &72   &84   &44.0   &3.90  &63 \\
HD~66811        &O4~I(n)fp     &220   &240  &107  &41.5   &3.70  &77 \\
HD~93204        &O5.5~V((fc))  &120   &105  &105  &41.0   &3.90  &47 \\
HD~93843        &O5~III((fc))  &90    &90   &40   &39.0   &3.65  &60 \\
{CD$-$47\,4551} &O5~Ifc        &100   &50   &110  &37.5   &3.55  &63 \\
{\it CPD$-$59\,2600}&O6~V((f)) &160   &120  &90   &40.5   &3.95  &40 \\ 
HD~169582       &O6~Iaf        &100   &73   &105  &37.0   &3.40  &80 \\ 
{\it HD~63005}  &O6.5~IV((f))  &80    &63   &87   &38.5   &3.75  &45 \\
HD~91572        &O6.5~V((f))   &75    &49   &73   &39.0   &3.90  &38 \\
{\it CD\,$-$43\,4690}&O6.5 III &110   &93   &103  &37.0   &3.60  &53 \\
HD~69464        &O7~Ib(f)      &90    &83   &92   &36.0   &3.50  &55 \\
HD~94963        &O7 II(f)      &100   &82   &82   &36.0   &3.50  &58 \\
HD~93222        &O7~V((f))     &80    &52   &90   &38.0   &3.80  &40 \\
HD~91824        &O7 V((f)z     &70    &47   &67   &39.0   &3.90  &37 \\
{\it HD~94370}  &O7.5~IInn     &230   &185  &84   &36.0   &3.50  &58 \\
{\it CPD\,$-$58\,2620}&O7.5~Vz &60    &39   &59   &38.0   &3.90  &35 \\
HD~75211        &O8.5~II       &130   &145  &58   &34.0   &3.40  &55 \\
HD~151804       &O8~Iaf        &100   &67   &75   &30.0   &3.10  &80 \\
HD~97848        &O8~V          &60    &42   &74   &37.0   &3.90  &31 \\
{\it HD~302505} &O8.5~III      &80    &43   &65   &34.0   &3.60  &35 \\
{\it HD~92504}  &O8.5~V        &190   &155  &82   &35.0   &3.85  &28 \\
HD~148546       &O9~Iab        &105   &100  &95   &31.0   &3.20  &58 \\
HD~46202        &O9.5~V        &25    &15   &34   &34.0   &4.00  &22 \\ 
HD~152249       &OC9~Iab       &95    &65   &93   &31.0   &3.10  &80 \\
{\it HD~76968}  &O9.5 Ib       &83    &55   &62   &31.0   &3.20  &58 \\
{\it CD\,$-$44\,4865}&O9.7~III &80    &60   &79   &30.0   &3.30  &38 \\
{\it HD~152003} &O9.7 Iab      &90    &77   &80   &30.5   &3.05  &83 \\
{\it HD~75222}  &O9.7 Iab      &90    &67   &80   &30.0   &3.10  &65 \\  
{\it HD~78344}  &O9.7 Iab      &80    &64   &64   &30.0   &3.00  &80 \\
{\it HD~69106}  &O9.7~III      &320   &310  &105  &30.0   &3.50  &27 \\ 
\hline
\end{tabular}
\end{center}
\small
{\bf Notes}: Spectral types are taken from \citet{sota} and  Sana et al. 
(in preparation) or are our own determinations (the stars highlighted by 
italics). Projected rotational velocities that do (\vsini\,(FT+GOF)) and 
do not (\vsini\,(PF)) account for extra broadening; \eb -- extra 
broadening velocities derived by assuming a radial-tangential distribution 
of photospheric turbulence; \Minit -- initial evolutionary masses  
determined from the tracks by \citet{brott}, with initial rotational 
velocities of $\sim$300~\kms. Uncertainties for \Minit:  +25/-37\%   for the 
higher and +12/-20\% for the lower mass end, respectively; for \Teff: 
$\pm$500 to 1500~K, and for \logg: +0.2/-0.1~dex. 
\end{table*}

\normalsize
While the number of Galactic B-type stars with reliably determined physical 
properties, including chemical abundances and  rotational and extra broadening 
velocities, has progressively increased during the past decade  (e.g., 
\citealt{morel06, morel08, dufton06, crowther06, hunter08, hunter09, MP, 
lefever10, fraser10}), O-type stars are currently under-represented in the 
stellar samples investigated so far (e.g., \citealt{bouret12, martins12a, 
martins12b}). 

Motivated by the lack of high-quality observational data, an extensive and 
detailed  spectroscopic survey of O and early-B stars in our Galaxy has been 
initiated by us and is currently under way. The immediate aim is to obtain 
reliable and coherent estimates for the photospheric and wind parameters 
(including nitrogen and helium abundances) and projected rotational rates for 
a statistically significant number of stars while allowing for the effect of 
extra broadening. By means of these data and incorporating similar data from 
the literature, we plan (i) to address the important question on the nature 
and the origin of extra broadening in the spectra  of  hot massive stars and 
(ii) to test model predictions about the evolution of rotation  and the role 
of rotational mixing in surface chemical enrichment during the main-sequence 
(MS) phase. In this paper, we present the main results from our analysis of 
31 O-type stars from the southern hemisphere. Results from the analysis of an 
even larger sample, observed within the IACOB project \citep{simon11}, will be 
presented in a forthcoming study (Sim\'on-D\'iaz et al., in prep.).
 
Our paper is structured as follows: in Section~\ref {observation} we describe 
the observational material and its reduction. In Section~\ref{methods} we 
outline the procedures used to determine the fundamental stellar parameters 
and to disentangle and measure the relative contributions of rotational and 
extra broadening. The main results are outlined and discussed in Sects~\ref{Ostars}, 
and ~\ref{OBstars} and are summarized in Sect.~\ref{conclusions}.

\section{Stellar sample and observations}\label{observation}

Our basic sample consists of 31  O-type stars in the Milky Way (MW),  selected 
by means of the following criteria: (i) to sample the luminosity classes I, 
III and V, and the spectral subtypes from O3 to O9.7; (ii) to be presumably 
single\footnote{Detailed information about the various criteria used to perform 
the multiplicity analysis can be found in Markova et al. (2011, Paper I), and in 
a forthcoming study (Markova et al., in preparation).}; (iii) to cover a wide 
range of projected rotational velocities; (iv) to be preferentially members of 
clusters and associations (i.e., to have known distances), and (v) to be bright 
enough ($V\le$10 mag) to allow for good quality spectra to be obtained during 
relatively short exposure times using medium-class telescopes. 

For 24 of the 31 targets we used our own observations, collected with  the FEROS 
spectrograph \citep{kaufer99} at the 2.2\,m MPI/ESO telescope at La Silla; for 
the rest, MPI-FEROS spectra available from the ESO archive, were used. All spectra 
have a spectral resolving power of 48\,000, with a typical signal-to-noise ratio 
(S/N) of about 200. One-dimensional, wavelength-calibrated spectra were extracted 
using the FEROS pipeline.  Stellar IDs  along with some spectral and physical 
parameters derived in the present study are listed in  Table~\ref{sample}.

\section{Spectral analysis}\label{methods}
\subsection{Spectral classification}\label{SPT}

Because earlier classifications of O-type stars can be subject to significant 
uncertainties (see, e.g., \citealt{markova11} and \citealt{sota}), we refrained 
from using these classifications, but proceeded as follows: for the stars in 
common with \citet{sota} and Sana et al. (in preparation), we adopted the 
spectral types and luminosity classes as provided in the corresponding studies; 
for the rest (highlighted by italics in Column~1 of Table~\ref{sample}) we derived 
our own classifications, following the premises by \citet{sota} and consulting 
the atlas of Galactic O-type standards, as compiled by Sana et al. Spectral types 
and luminosity classes of our sample stars are listed in Column 2 of Table~\ref{sample}.  
The accuracy of these data is typically one subtype.

\subsection{Measuring the projected rotational and extra broadening velocities} 
\label{vrot_vmac}

Projected rotational velocities (\vsini) of OB stars were estimated by applying 
various methods, among which the most commonly used are: the full-width-at-half-maximum  
(FWHM) method  (e.g., \citealt{abt02}); the goodness-of-fit (GOF) method (e.g., 
\citealt{CE77, ryans02});  the cross-correlation (CC) method (e.g., \citealt{penny96, 
howarth97}), and the Fourier transform (FT) method (viz \citealt{gray76, gray05}, see 
also \citealt{simon07, MP, fraser10}). While the CC and the FWHM methods work under 
the assumption that factors other than rotation do not significantly affect the line 
widths, the GOF and the FT methods are, in principle, able to separate rotational 
broadening  from other broadening mechanisms.
 
To separate and measure the relative contributions of rotational and extra broadening 
in our O-star sample, we considered a newly developed FT+GOF approach and applied it 
to the \OIII line\footnote{For three stars with very weak \OIII (CD\,$-$44\,4865, 
HD~78344 and HD~69106), \SiIIIa or N~III $\lambda$4515  absorption lines  were used 
instead.}. The  method consists of a three-step procedure: $first$, the FT of the line 
is computed, and from the position of the first zero the corresponding projected rotational 
velocity (\vsini(FT)) is  estimated; $second$, the observed profile is compared with 
a synthetic one, degraded to the resolving power of the analyzed spectrum, and convolved 
with broadening functions calculated for an appropriate range of projected rotational 
and extra broadening rates. From the minimum in the $\chi^{2}$ distribution, optimum 
values for the two broadening agents  (\vsini(GOF) and $\Theta$(GOF)), are derived. 
$Third$, by  a thorough analysis of all estimates thus obtained, the final values of 
the rotational and extra broadening rates are fixed. For the particular case of our 
sample, we adopted as final values the \vsini\, derived with the FT method and 
the GOF solution for extra broadening -- obtained by assuming a radial-tangential (RT) 
distribution of velocities with a radial component being equal to the tangential one 
(hereafter \eb)\footnote{A detailed description of the FT+GOF method along with a 
discussion of its applicability and limitations in the case of O and early B-type 
stars can be found in \citet{SH13}.}.
\begin{figure*}
\begin{minipage}{5.9cm}
\resizebox{\hsize}{!}
{\includegraphics{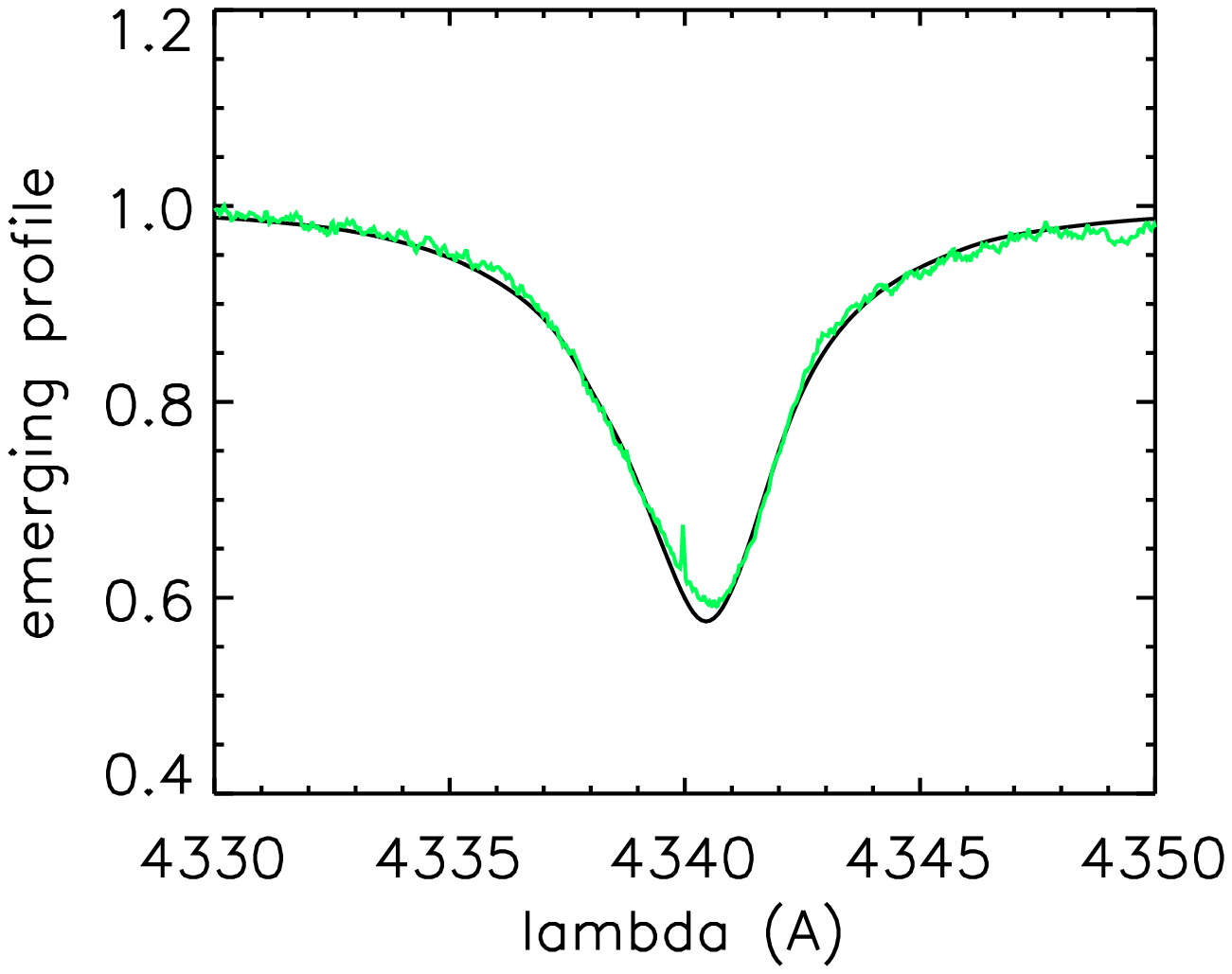}}
\end{minipage}
\hfill
\begin{minipage}{5.9cm}
\resizebox{\hsize}{!}
{\includegraphics{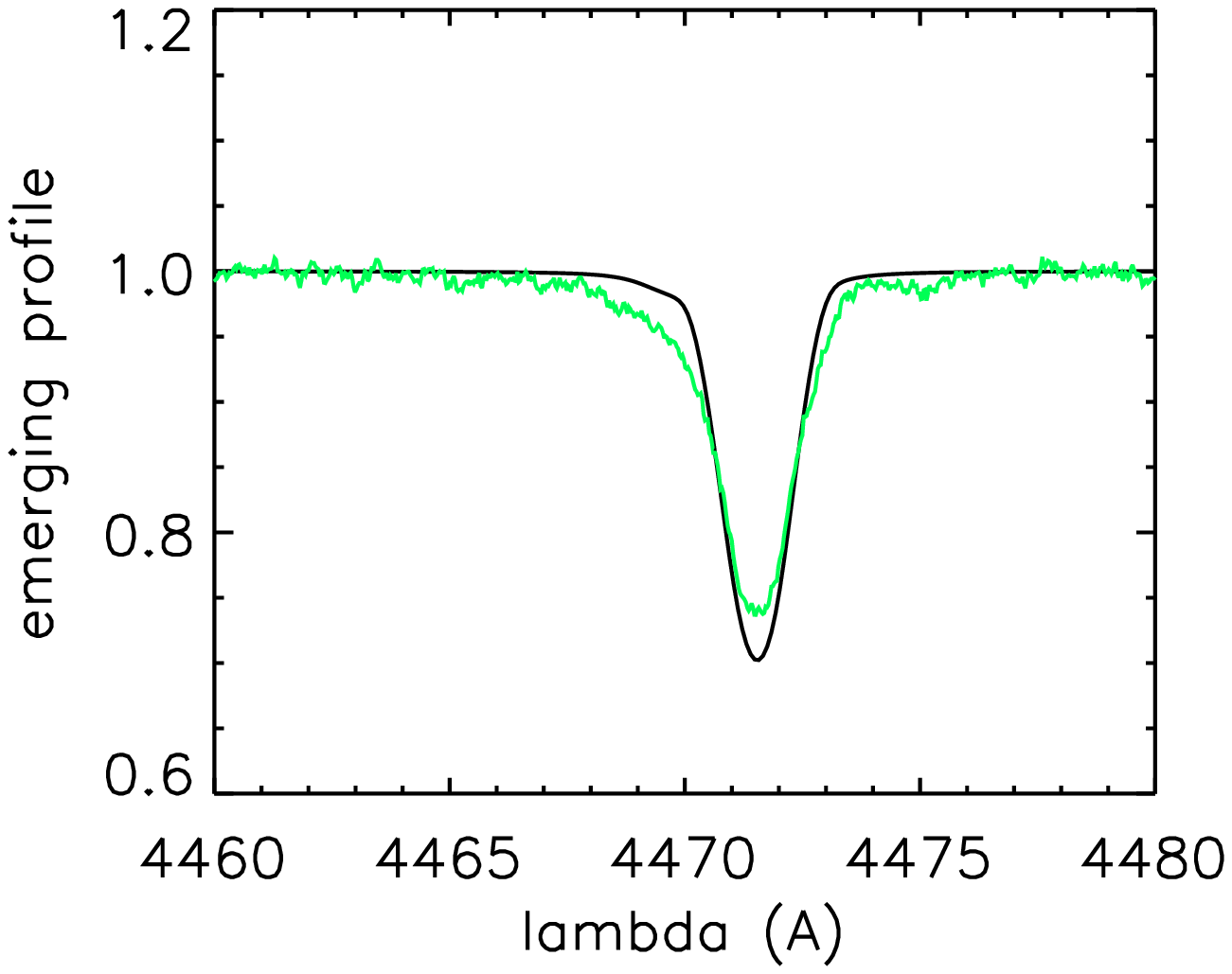}}
\end{minipage}
\hfill
\begin{minipage}{5.7cm}
\resizebox{\hsize}{!}
{\includegraphics{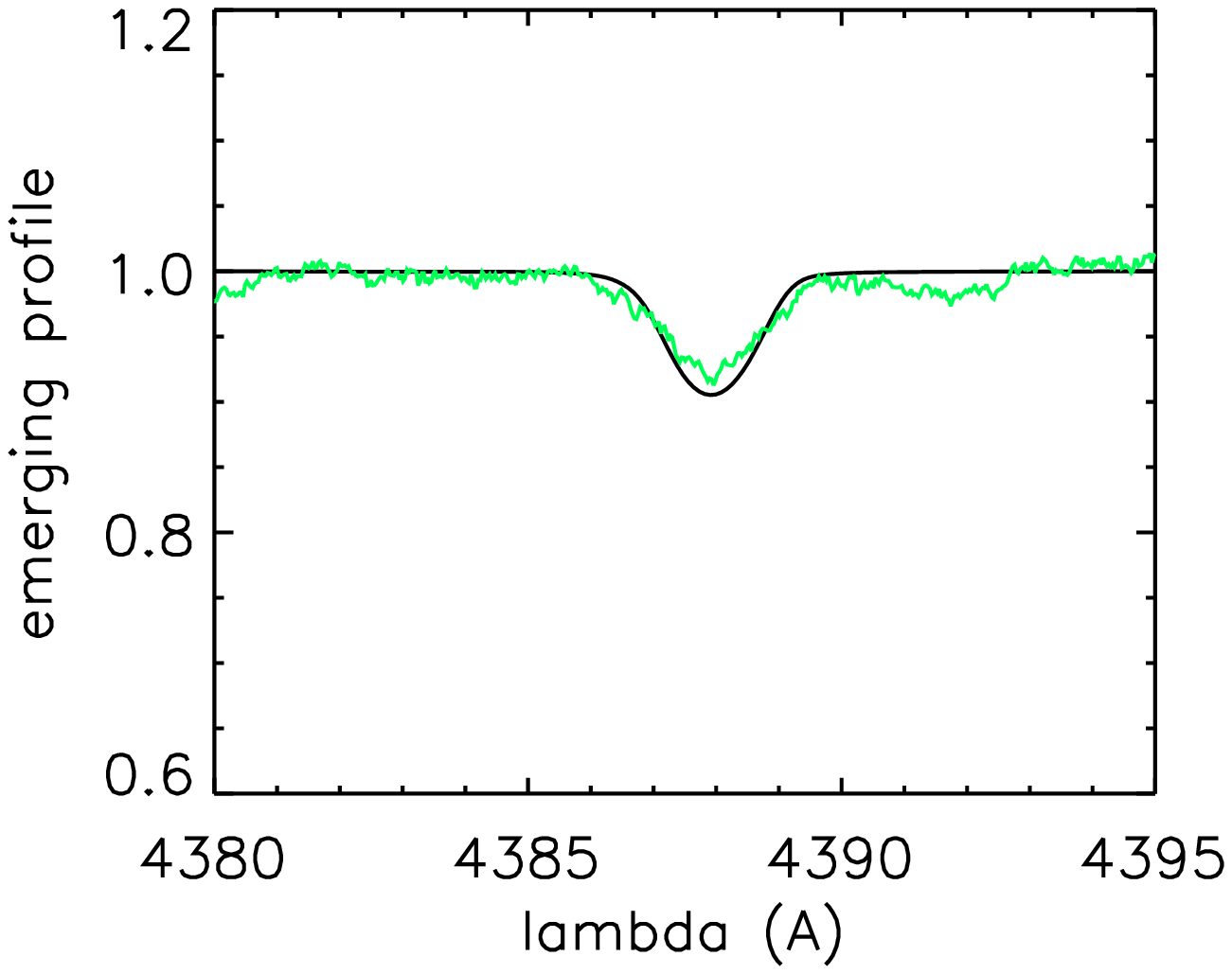}}
\end{minipage}
\\
\begin{minipage}{5.9cm}
\resizebox{\hsize}{!}
{\includegraphics{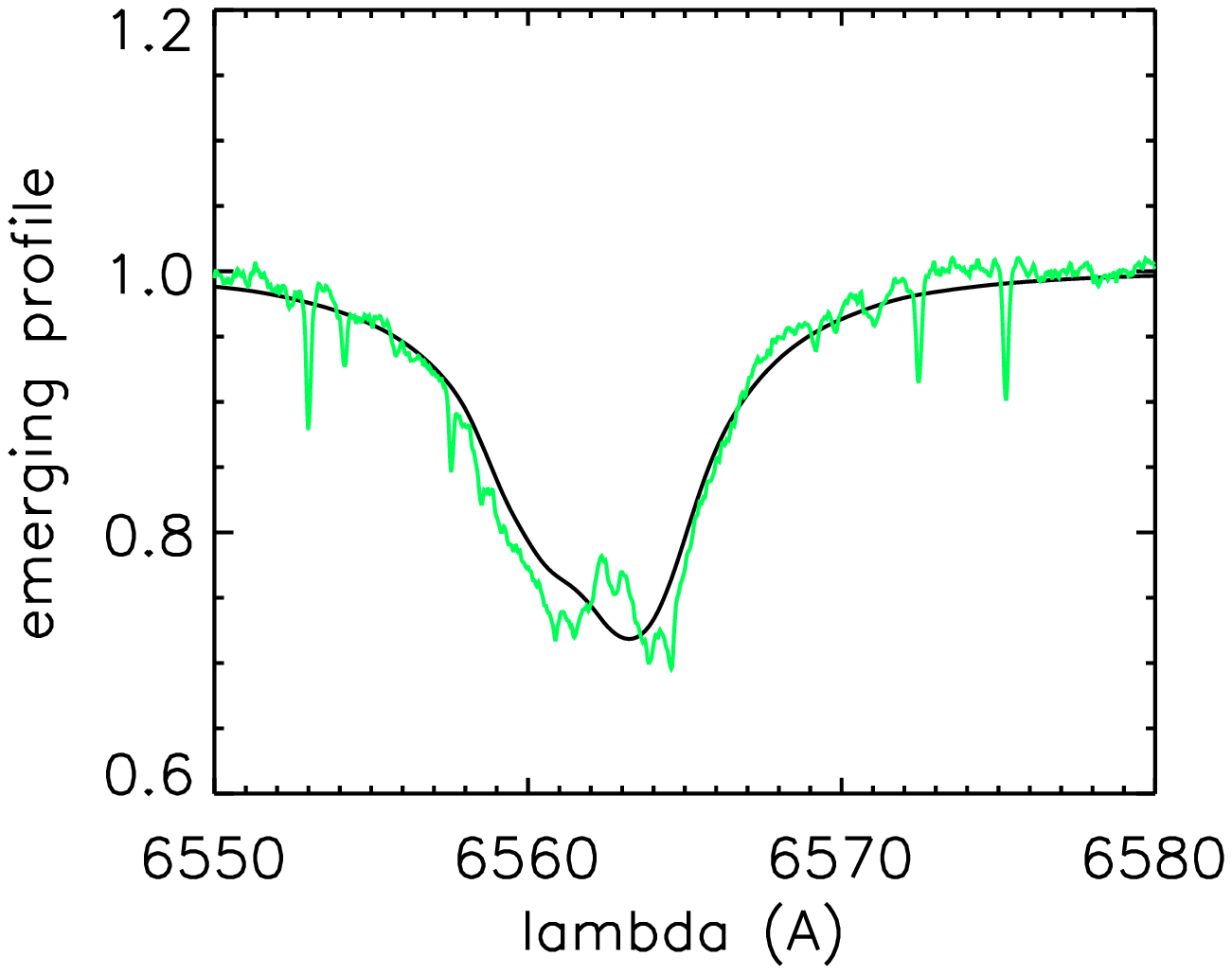}}
\end{minipage}
\hfill
\begin{minipage}{5.7cm}
\resizebox{\hsize}{!}
{\includegraphics{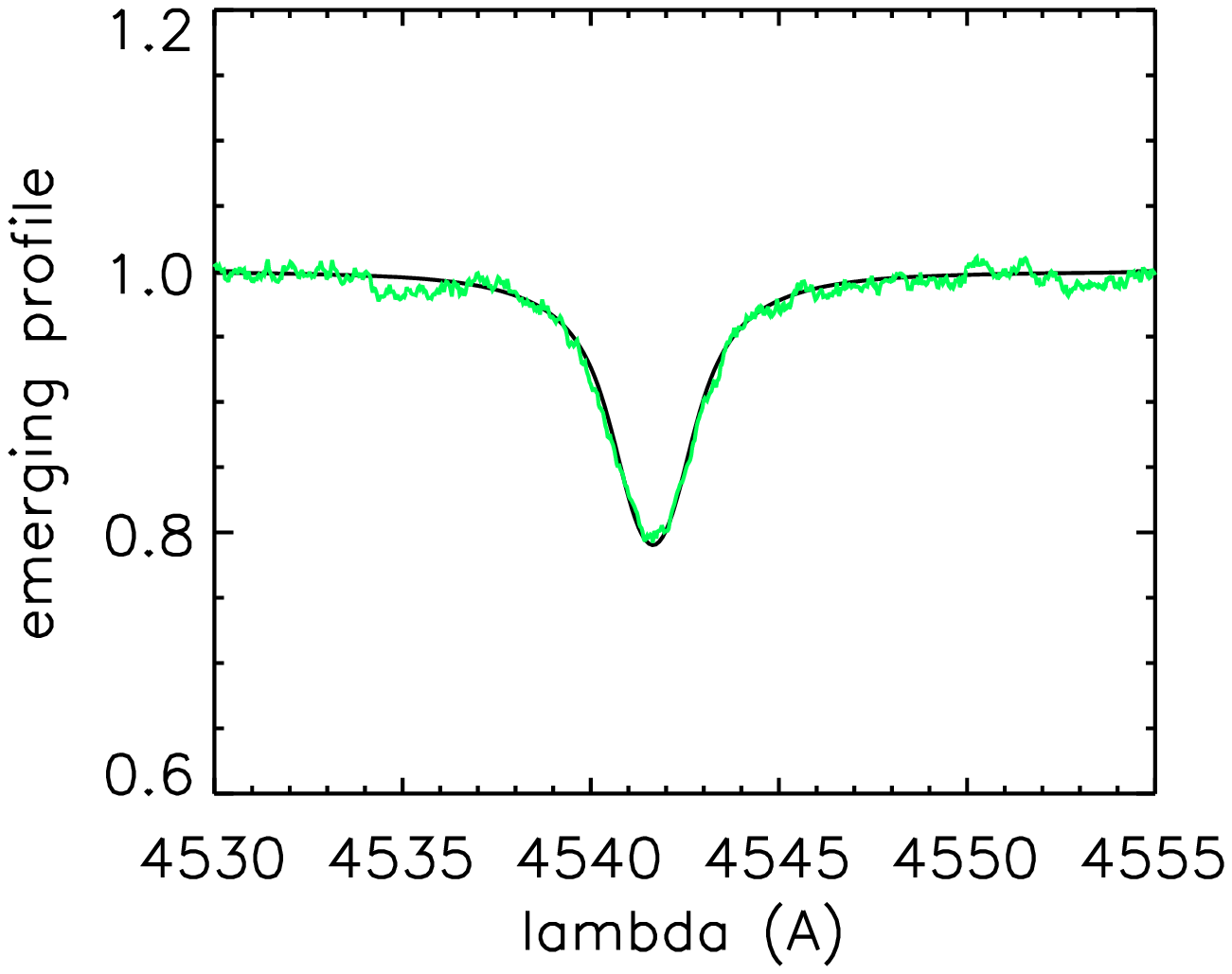}}
\end{minipage}
\hfill
\begin{minipage}{5.7cm}
\resizebox{\hsize}{!}
{\includegraphics{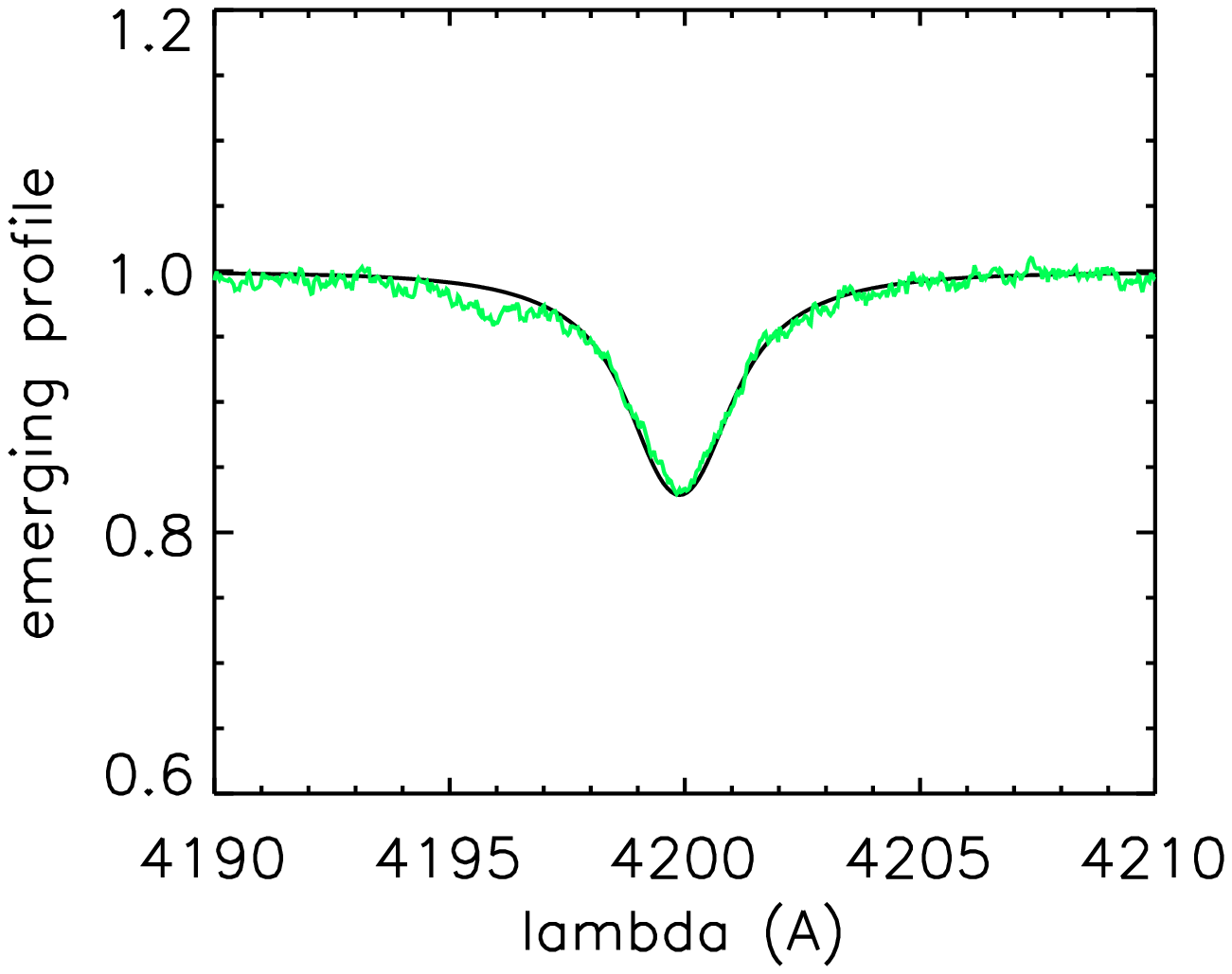}}
\end{minipage}
\caption{Illustration of the quality of the fit to strategic helium and 
hydrogen lines for  a star with a weak wind, HD~93222.
}
\label{figure_1}
\end{figure*}

The projected rotational and extra line-broadening velocities, obtained as 
described above,  are listed in Columns  4 and 5 of Table~\ref{sample}, 
respectively. The uncertainties in these estimates range from 10\% to 20\%, 
with higher values typical for more rapid rotators. The limit to the lowest 
\vsini, imposed by the  resolving power of the spectrograph, equals  $\sim$12~\kms. 
Since the quality of our spectra is relatively high, while the \OIII\, line 
used in this analysis is relatively weak, we do not expect our \vsini\, to be 
significantly affected by instrumental effects, as described in \citet{simon07}. 
On the other hand, for stars with fast rotation ($v_{rot}\ge$~200~\kms), 
processes such as differential rotation, limb or gravity darkening may be 
present, which will alter the profile shape and thus its Fourier transform 
(see, e.g., \citealt{reiners03}), while  for those with slow rotation  
microturbulence may be a problem  \citep{gray73}. These possibilities 
are taken into account when we discuss the outliers (if present). 

Finally,  we point out here that we are aware that  our description of extra 
broadening in terms of large-scale turbulent motions may be problematic (because 
of the obtained highly supersonic velocities), and that an explanation in 
terms of collective effects from numerous high-order, low-amplitude nonradial 
g-mode pulsations has been suggested by \citet{aerts09}. Nevertheless, we 
consider our approach  suitable because (i) it allows to compare our results 
for O-type stars with similar results for B-type stars from the literature  
(see Sect.~\ref{OBstars}), and (ii) the pulsation hypothesis, although promising 
for massive early B-type supergiants \citep{simon10}, needs to be confirmed in 
a wider parameter space. On the other hand,  {\it if} stellar pulsations were 
responsible for the extra broadening, a risk of  underestimating the actual 
\vsini\, when applying the FT method may be present \citep{aerts09}.

\subsection{Atmospheric parameters}
\label{stellar_par}

Stellar and wind parameters of our basic O-star sample were  determined  from the best 
fit (estimated by eye) between observed and theoretical profiles of strategic lines. 
The synthetic profiles were calculated by applying a recently updated version of 
the FASTWIND code (see \citealt{santolaya} and \citealt{puls05} for previous versions, 
and \citealt{gonzalez12} for a brief summary of the latest version), and  were 
broadened using corresponding values of  \vsini\, and \eb\, as listed in Columns 
4 and 5 of Table~\ref{sample}. Detailed information about the model atmosphere 
analysis of our sample stars will be presented in a forthcoming paper (Markova et al., 
in preparation), while here we note that all models were calculated by adopting  
a microturbulent velocity of 10~\kms for both the atmospheric structure calculations 
and the line profile synthesis. By default, normal helium and nitrogen abundances 
(\Yhe = $N_{\rm He}/N_{\rm H}$ = 0.10 and [N]~=~log N/H +12 = 7.7) were used 
and were adjusted when necessary. An example of a good-quality fit is provided in Fig.~\ref{figure_1}.

The resulting values for \Teff\, and \logg\, are listed in Co\-lumns 6 and 7 of 
Table~\ref{sample}, respectively. The error on these data, estimated from the best 
fit to the strategic lines, is typically $\pm$500 to 1500~K in \Teff\, and  
$\pm$0.1~dex in \logg\, (but see Sect.~\ref{consist_check}). 

\begin{figure*}
{\includegraphics[width=5.7cm,height=5.cm]{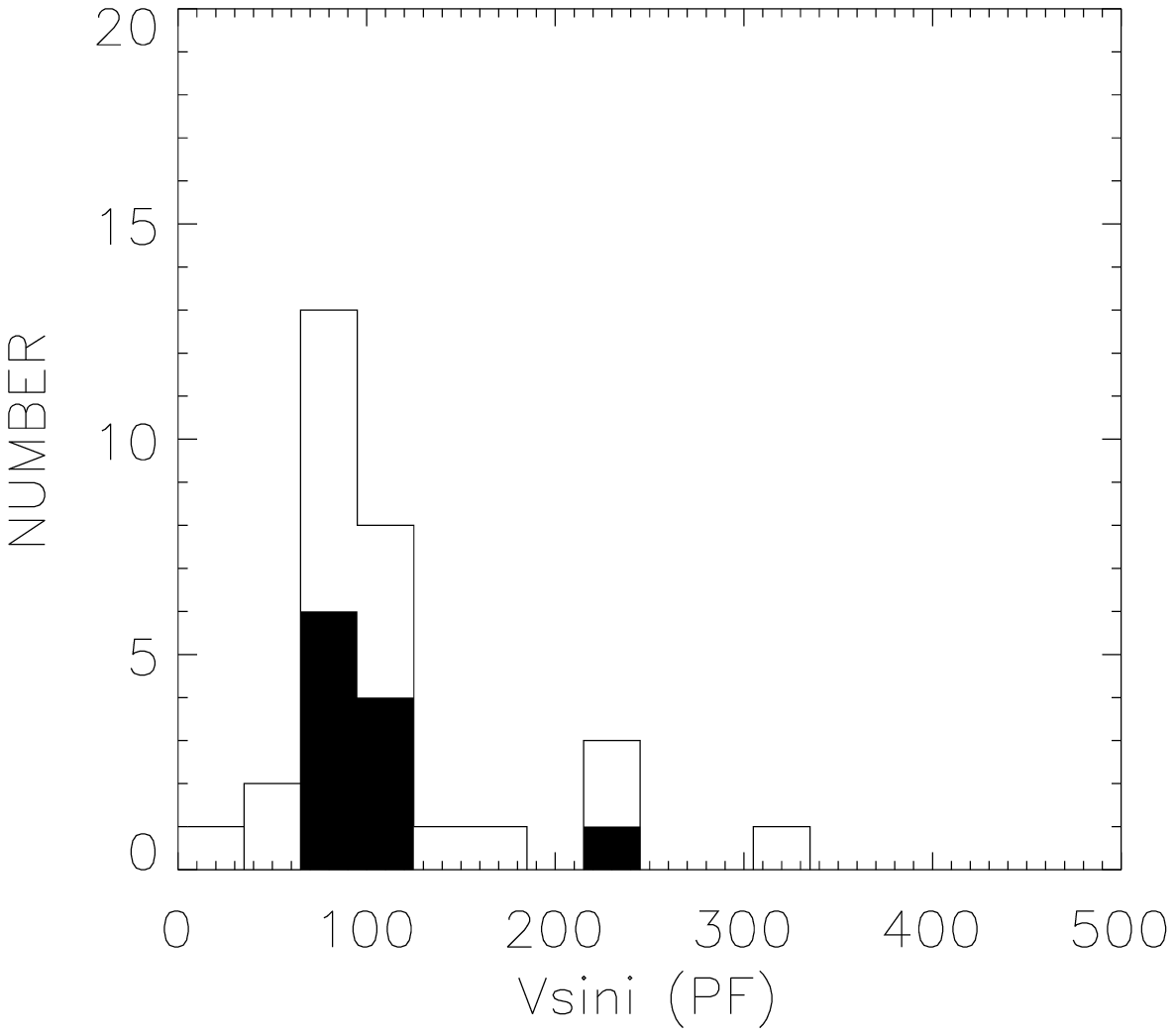}}
{\includegraphics[width=5.7cm,height=5.cm]{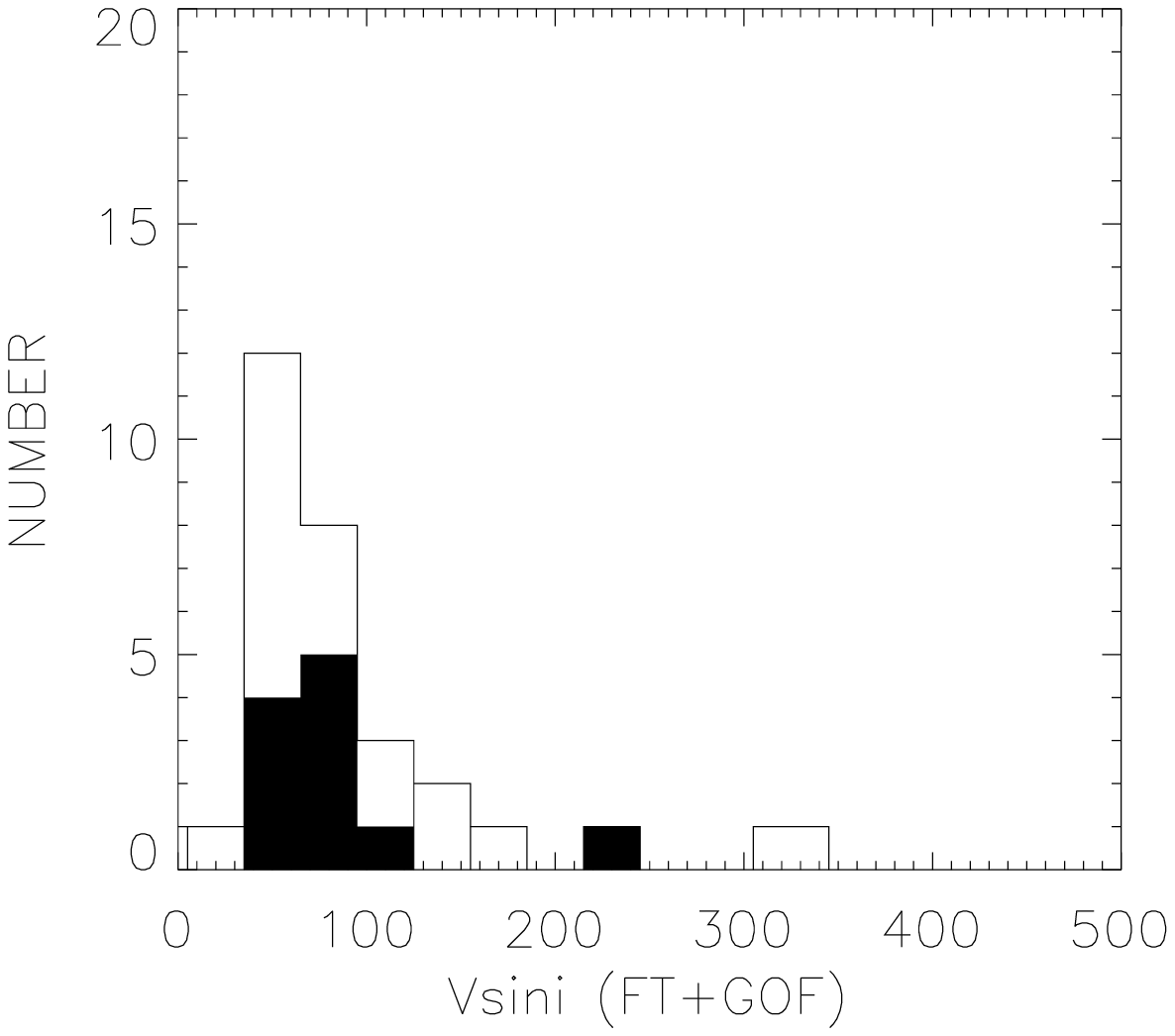}}
{\includegraphics[width=6.7cm,height=5.cm]{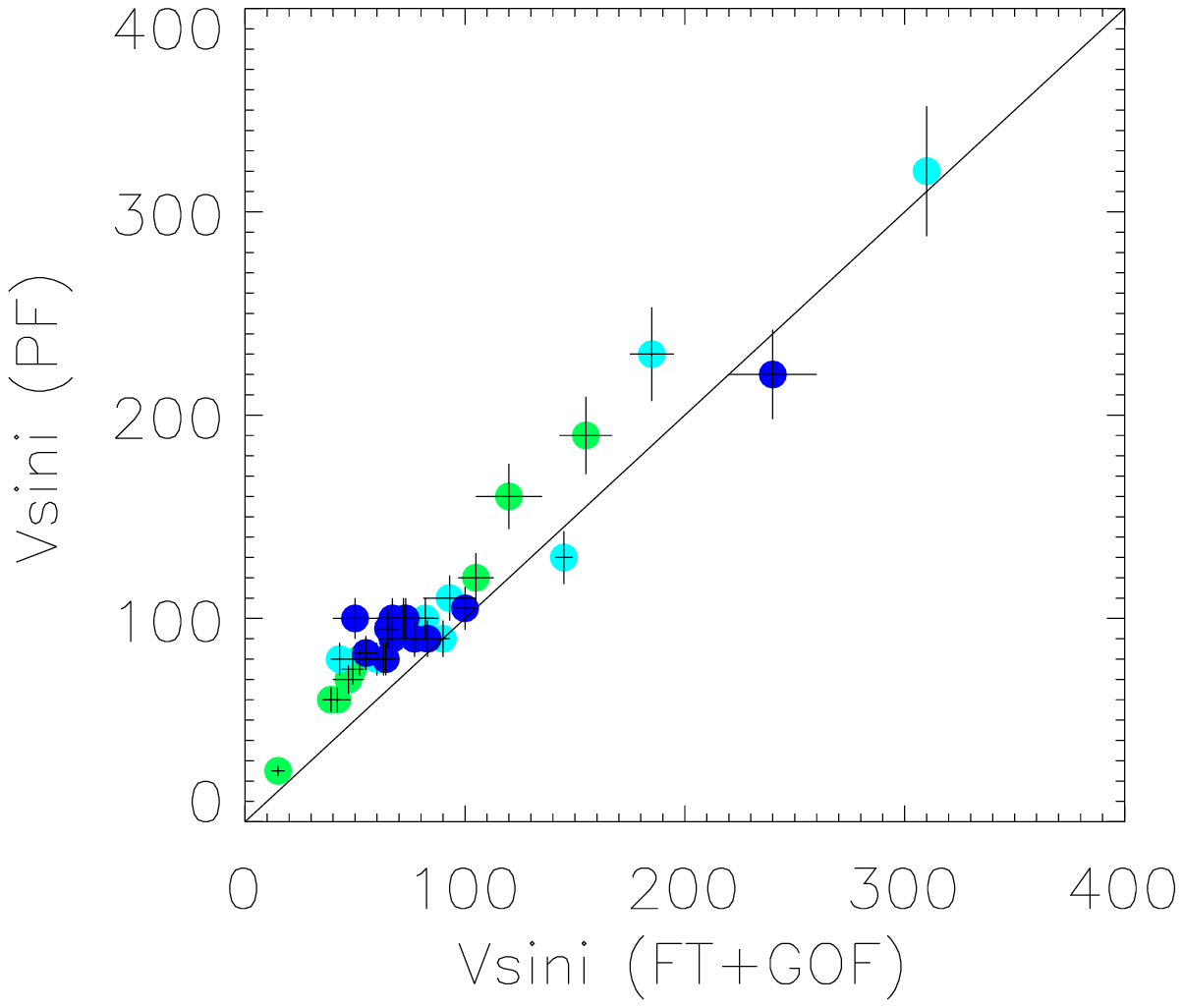}}
\caption{{\it Left and middle panel}: Distributions of projected rotational 
velocities for the O-star sample, based on our determinations for \vsini\,(PF) 
(extra line-broadening neglected) and \vsini\,(FT+GOF) (extra line-broadening 
taken into account). In both panels, the binsize is 30~\kms, and the contribution 
of supergiants is highlighted by the shaded area. \newline {\it Right panel}: 
\vsini\,(FT+GOF)  versus \vsini\,(PF). Luminosity class I objects are highlighted 
in dark blue, II and III objects in light blue, and IV and V objects in green. 
The one-to-one relation is marked by the solid line. 
}
\label{figure_2}
\end{figure*}

\section{Results for  the ESO O-star sample}\label{Ostars}

\subsection{Influence of extra broadening on the 
projected rotational velocities}\label{vrot_oldnew}

To reveal whether and to which extent the spectra of our sample O stars are 
influenced by effects of extra broadening,  we proceeded as  follows: $first$, 
we created an internally consistent dataset of \vsini\, without accounting for 
extra broadening (these data are referred to as \vsini(PF), see below); $second$, 
we analyzed these data for possible selection effects and effects caused by a 
limited number of objects, and $third$, we compared  the \vsini\,(PF) and the 
\vsini(FT+GOF) values, attributing the corresponding differences to the effect 
of extra broadening. 

\paragraph{Step 1}
For the majority of our sample stars, previous determinations of \vsini\, are 
available in the literature, mainly from \citet{penny96} and \citet{howarth97} 
but also from \citet{UF}, none of which provide a complete overlap. To avoid 
possible inconsistencies caused by the use of data from different sources, we 
decided to perform our own determinations of \vsini, which do not take into 
account extra broadening. To this end, for each sample star we used the best fit 
FASTWIND model, obtained as described in Sect.~\ref{stellar_par}, and optimized 
the fit between observed and synthetic lines now neglecting the extra broadening. 
The \vsini\, obtained in this way are listed in Column~3 of Table~\ref{sample} 
and are referred to as \vsini(PF),  where PF abbreviates {\it profile fitting 
method}. The uncertainty in these data, determined by the quality of the fits 
estimated by eye, is typically lower than 10~percent.   

\paragraph{Step 2}
In their studies of statistically significant samples of Galactic O stars, 
\citet{penny96} and \citet{howarth97} found that the derived distributions 
of \vsini\, (which do not account for extra broadening!) are quite peculiar, 
showing  a low-velocity maximum at about 80~\kms and a high-velocity tail 
dominated by MS stars. Supergiants show fewer rapid rotators (\vsini$\ge$200~\kms), 
and they also lack sharp-lined stars  (\vsini$\le$60~\kms). 

An inspection of the data shown in the left panel of Fig.~\ref{figure_2} indicates 
that the \vsini\,(PF) distribution of our basic O-star sample  is characterized by 
features similar to those described by \citet{penny96} and \citet{howarth97}.
Comparing this distribution and the one derived for the case when extra 
broadening is accounted for (middle panel of the same figure), we furthermore 
note that while in general the stars move to lower velocities, the sample still 
lacks slowest projected rotators and may also lack fast rotators. Taken at 
face value, these findings might indicate that either the sample is biased 
toward objects with only medium high and medium low velocities, or that in the 
parameter space covered by our targets very fast and very slow rotators are
intrinsically rare.

\paragraph{Step 3}
The right panel of Fig.~\ref{figure_2}  shows the correspondence between 
\vsini(PF) and \vsini(FT+GOF). Apparently, for the majority of sample stars 
the velocities originating from the combined FT+GOF approach are lower than 
those obtained via the profile-fitting method. The differences are larger 
than the error budget of the two datasets, and  we ascribe them to the effect 
of extra broadening. In agreement with previous results from \citet{simon11}, 
we see that not only supergiants but also O-type giants and dwarfs are subject 
to significant extra broadening.  The influence of this broadening on the 
projected rotational velocities, averaged over the whole sample, is estimated 
to be -20$\pm$15~\kms, in perfect agreement with similar results from 
\citet{bouret12} regarding supergiants. 

\subsection{Extra broadening velocities}
\label{vmac_ostars}

\begin{figure}
{\includegraphics[width=8.cm,height=5.5cm]{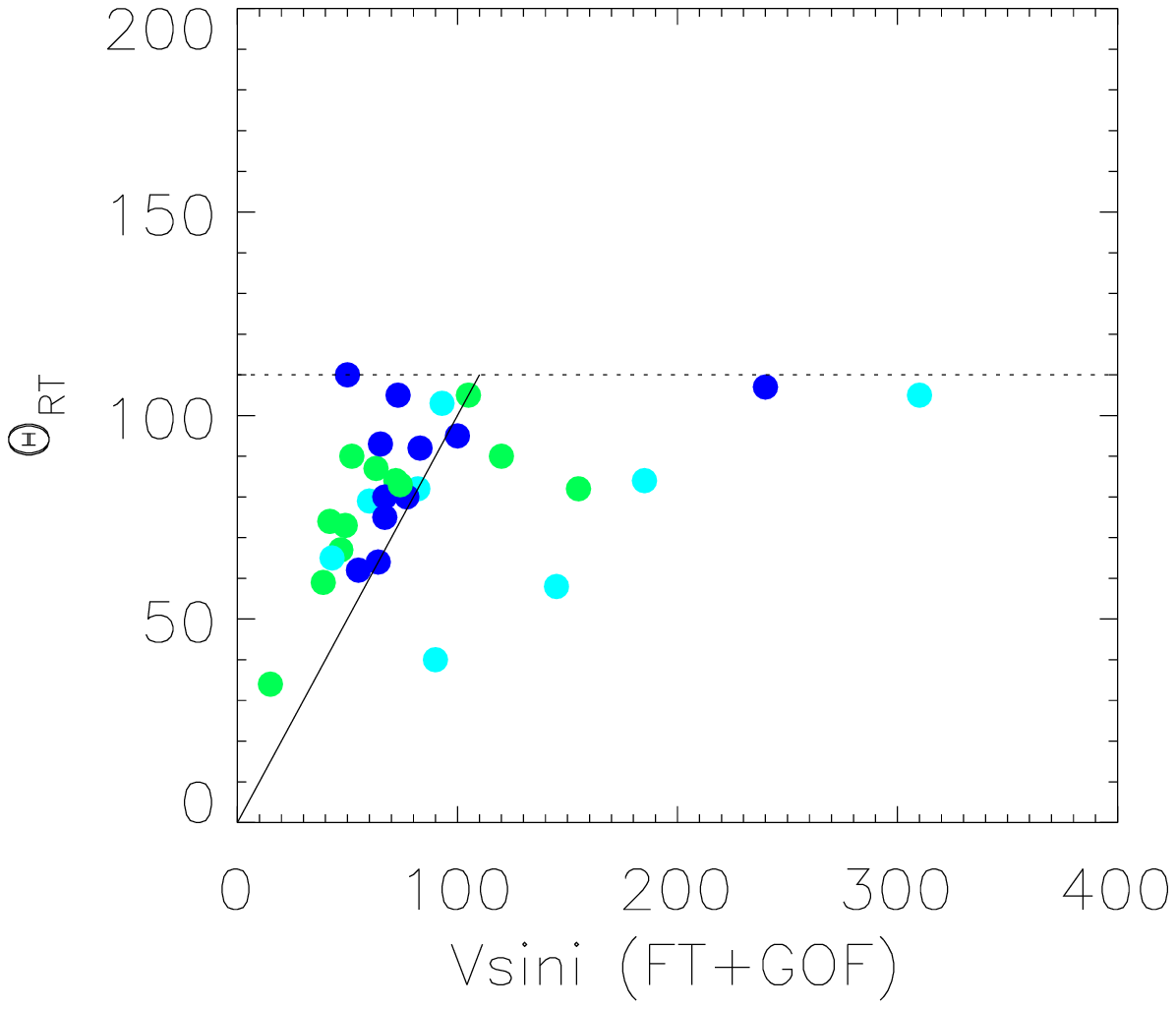}}
\\
{\includegraphics[width=8.cm,height=5.5cm]{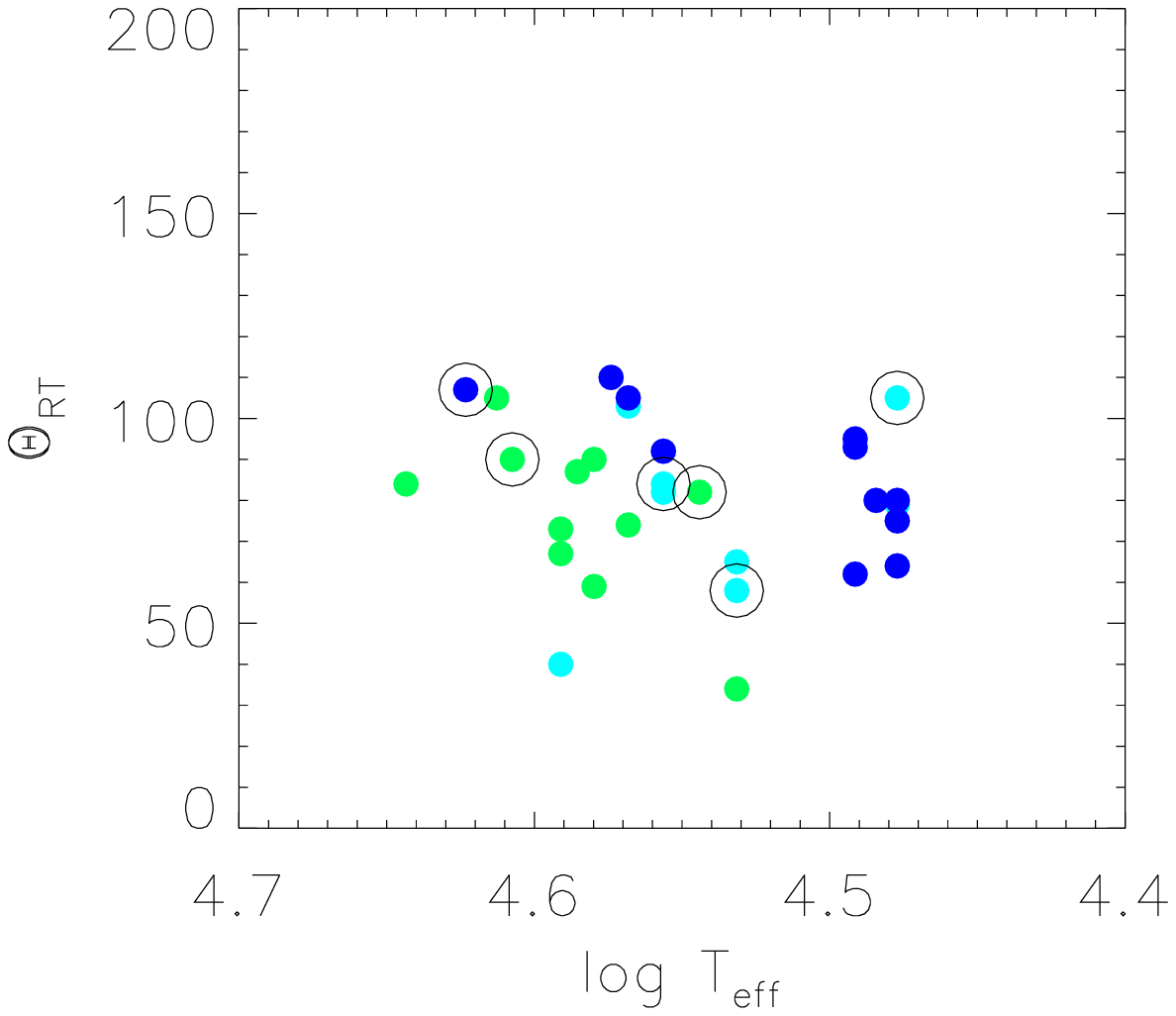}}
\caption{Extra broadening rates (in \kms) for the O-star sample as a function 
of \vsini\, (upper panel) and \Teff\, (lower panel). In both panels, luminosity 
class I stars are colored in dark blue, luminosity class II and III objects in 
light blue,  and luminosity class IV and V objects in green. Intermediate 
(110$\le$\vsini(FT+GOF)$\le$200~\kms) and fast rotating stars 
(\vsini(FT+GOF)$>$200~\kms) are additionally marked by large circles. 
In the upper panel, the solid line indicates the one-to-one relation.} 
\label{figure_3}
\end{figure}

Fig.~\ref{figure_3} depicts the  \eb\,values as a function of  \vsini\footnote{Since 
it is not clear in advance whether the additional broadening is subject to 
projection effects or not, here and in the following we concentrated on projected 
rotational velocities.} 
(upper panel) and \Teff\, (lower panel). Different luminosity classes are 
colored differently. Intermediate and fast rotators\footnote{For the purpose 
of this analysis, $intermediate$ and $fast$ rotating O stars refers 
to objects with 110$\le$\vsini$\le$200~\kms and \vsini$>$200~\kms, respectively.}  
are additionally marked by large circles.

From the upper panel of this figure, we find that the sample can be divided 
into two groups: objects in which rotational broadening is stronger than  
extra broadening, and those in which extra broadening either dominates or is 
in strong competition with rotation. The majority of stars appears to be 
members of the second group. Interestingly, and on a general scale as well as 
within each of the two groups, the \eb\,values tend to increase toward higher 
\vsini, without exceeding  a specific value of 110~\kms, however. Marginal 
evidence of supergiants with  higher extra broadening velocities than dwarfs 
and giants seems to present as well: the mean values of \eb\, averaged over the
subsamples of supergiants, giants, and dwarfs equals  88$\pm$17,
77$\pm$22 and 76$\pm$19~\kms, respectively.

From the lower panel we furthermore note that while within the total 
sample no clear indication of a connection between \eb\, and \Teff\, can be 
seen, within the subgroups of supergiants and dwarfs a tendency 
for the hotter objects to exhibit higher \eb\, may be present. This
possibility has been confirmed by a Spearman rank-correlation test, 
providing a correlation coefficient and a two-sided significance of 
its deviations from zero\footnote{Small numbers indicate a high 
significance.} (given in brackets)  of 0.21 (0.26), 0.80 (0.0003) and  
0.52 (0.10) for the total sample, and the subsamples of supergiants and 
dwarfs, respectively. Our findings are consistent with 
similar results from \citet{simon11} who also suggested that the extra 
broadening rates of hot massive stars most likely increase when moving 
from early-B to early-O subtypes, with values that are systematically higher  
for supergiants than for giants and dwarfs.

Interestingly, for the sample giants and bright giants, we were unable   
to confirm the positive correlation between \eb\, and \Teff, as 
suggested by \citet{simon11}: for this subgroup of stars  we basically found 
no correlation at all (SR correlation coefficient with a two-sided 
significance of  $-0.22~(0.60)$). The explanation for this inconsistency
is currently unclear, but reasons such as a limited number of objects (eight 
in total), unrecognized binarity, and possible biases toward intermediate 
and fast rotators - almost half of these stars rotate with 
\vsini$\ge$110~\kms - might equally contribute. 
 
Summarizing our results, we conclude that the spectra of O stars (at least 
those of solar metallicity) are subject to significant extra broadening, 
whose effect (when expressed as a velocity) most likely scales with effective 
temperature and luminosity.

\section{Extending the sample toward B-supergiants}\label{OBstars}

The spectra of massive B-supergiants have been proven to experience 
significant extra broadening (e.g., \citealt{dufton06, lefever07, MP, 
fraser10, simon07, simon10}). To obtain more insight into the 
properties of this enigmatic phenomenon, we decided to extend our 
analysis toward lower temperatures and lower surface gravities, 
incorporating appropriate data for B supergiants from the literature. 

There are three Galactic studies \citep{lefever07, MP, fraser10} that 
investigate atmospheric parameters and rotational and extra broadening 
velocities of massive B-supergiants, applying methodologies similar to 
ours (but see next section). From these studies we selected 
73 targets with {\it non-negligible} extra broadening, 46 from 
\citet{fraser10}, 19 from \citet{lefever07} (their $group$~I stars only), 
and eight from \citet{MP}. To these, we added five more B-supergiants 
from the work of \citet{simon10}, for which reliable determinations of 
\Teff\, and \logg\,  were found in the literature (from \citealt{searil} 
and \citealt{markova08}). 

Recently, \citet{bouret12} have published stellar and wind parameters 
and projected rotational and extra line-broadening rates for a 
sample of eight massive O-supergiants in the MW, derived 
by means of a combined FUV, UV, and optical analysis. Since five of 
these are intermediate and fast rotators, while the number 
of such stars in our ESO O-star sample is limited to only two (see 
Fig.~\ref{figure_2}), we incorporated these eight objects into 
our analysis as well. Thus, the total number of objects in 
our enlarged OB sample amounts to  117,  39 of which are O stars and
78 are massive B-ype supergiants.

\subsection{Consistency check}\label{consist_check}  

Ideally, one would like to perform a consistency check by comparing 
stars in common between different datasets. Unfortunately, however, 
the number of such objects among the sample OB stars is too limited 
to allow for firm conclusions: there is one star in common between 
our O-star sample and the one studied by \citet{bouret12} (HD~66811);
%
%
%
three stars between \citet{simon10} and \citet{MP} (HD~190603, HD~206165 and 
HD~191243), and one between \citet{fraser10} and \citet{lefever07} 
(HD~148688). Nevertheless, it is encouraging  that in all these cases 
the corresponding estimates agree within the error. 

On the other hand, in three of the five studies mentioned above, 
namely \citet{lefever07}, \citet{MP}, and \citet{simon10},  
a methodology similar to ours has been used to separate and measure 
the relative contributions of rotation and extra broadening, suggesting 
that systematic differences between the corresponding datasets are
unlikely.
\begin{figure}
{\includegraphics[width=7.5cm,height=5.5cm]{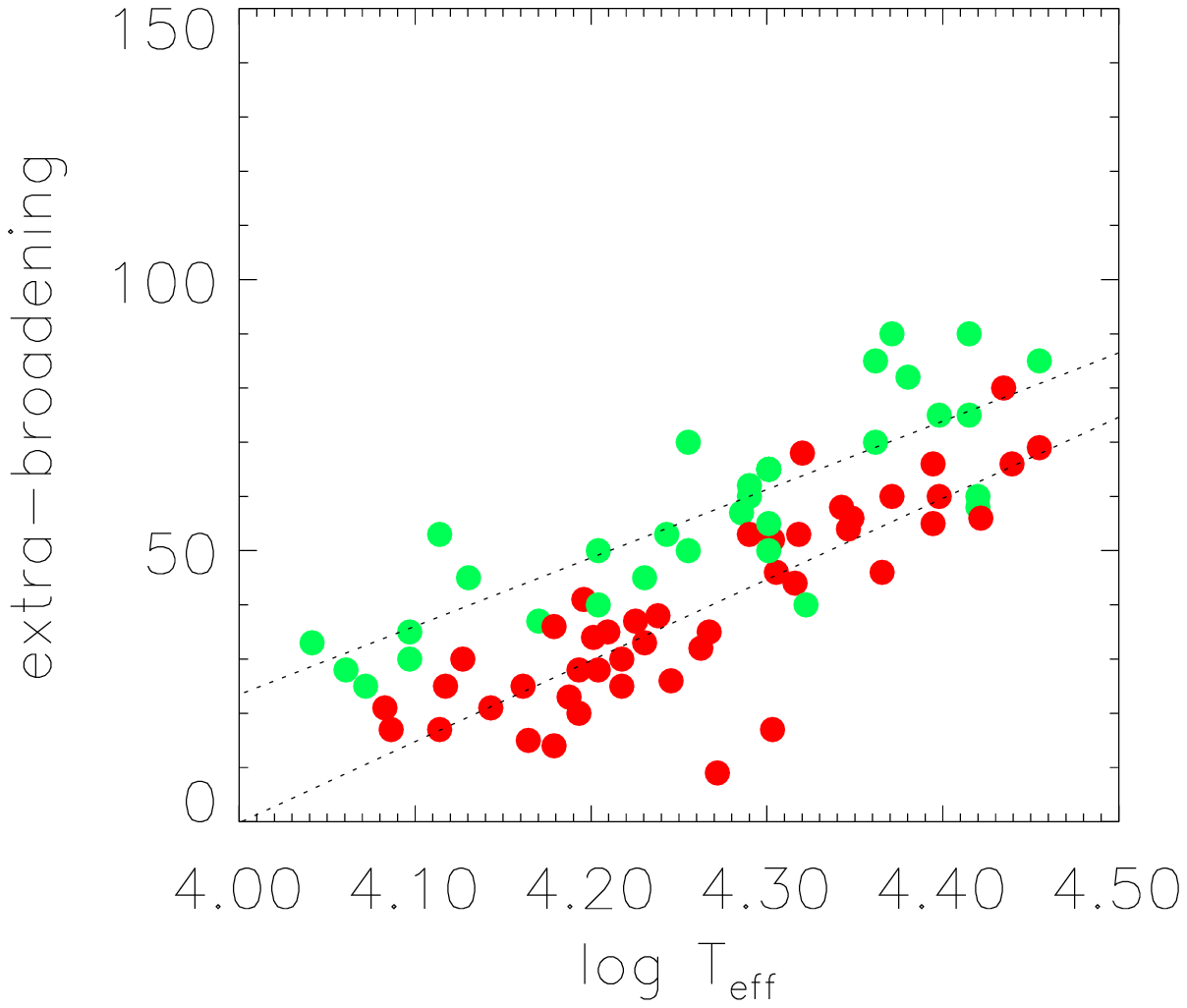}}
{\includegraphics[width=7.5cm,height=5.5cm]{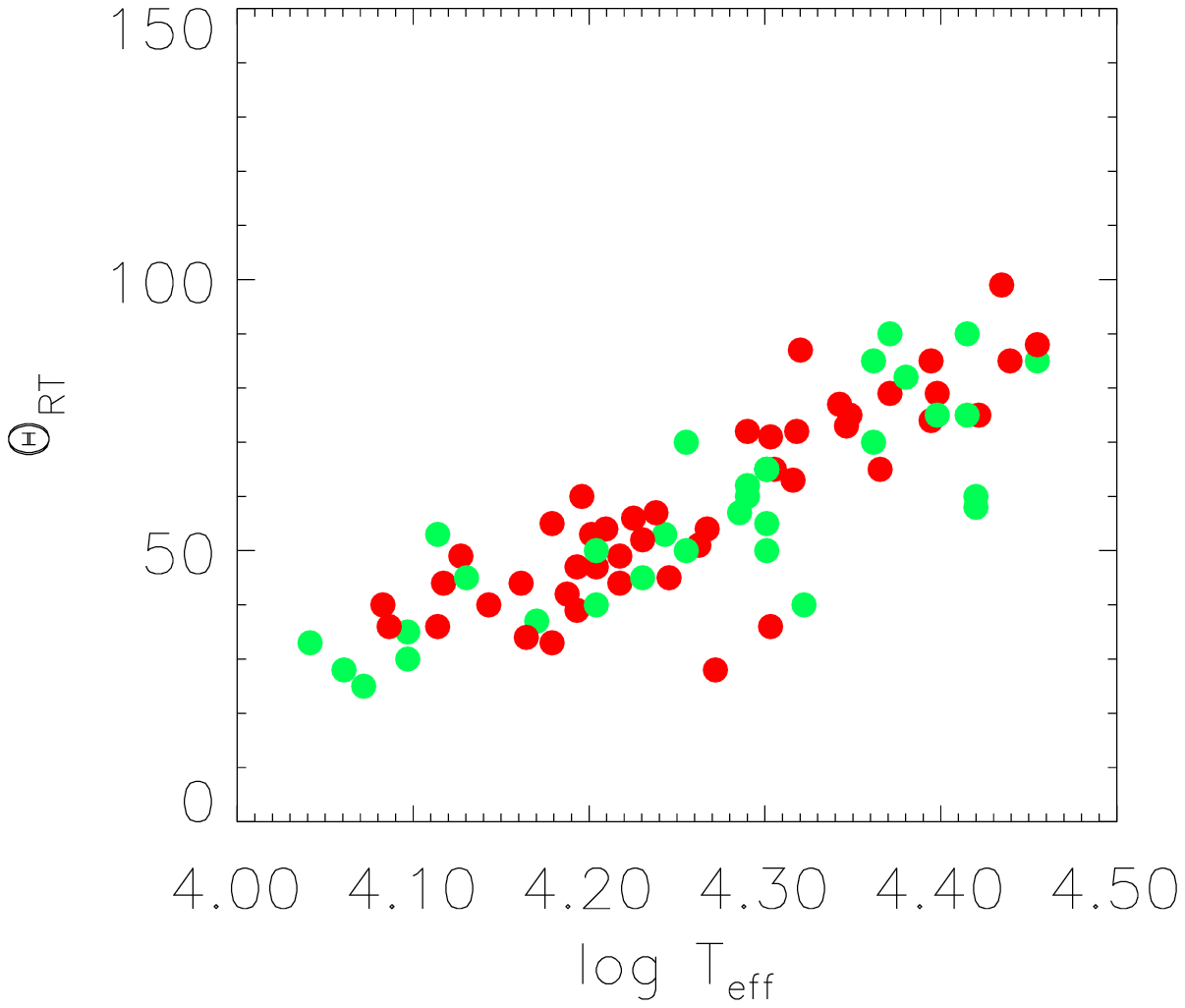}}
\caption{Distribution of extra broadening velocities for the 
B-supergiants in the OB sample. The data derived by assuming different 
models for extra broadening are highlighted in different colors: 
red denotes an isotropic Gaussian model \citep{fraser10}, whilst 
green refers to a radial-tangential one \citep{lefever07, MP, simon10}.  
Note the systematic shift along the vertical axes between the 
green and the red dots (upper panel) and that this shift disappears 
after the original Fraser et al. (2010) data for extra broadening 
have been corrected to make them consistent with the rest (lower panel). 
For dotted lines,  see Sect.~\ref{consist_check}}.
\label{figure_4}
\end{figure}

While in the investigations by \citet{fraser10} and \citet{bouret12} 
the \vsini\, values have been derived by applying the FT method and thus
are expected to be consistent with the rest of the OB sample, a
somewhat different approach has been adopted to quantify the amount of
extra broadening: instead of a radial-tangential model for the
turbulent velocity field, they used an isotropic Gaussian one. 

From a corresponding analysis of our ESO O-star sample, we derived that 
the RT model provides broadening velocities that are $\sim$20~\kms higher  
than those originating from an isotropic Gaussian model. The data shown 
in the upper panel of Fig.~\ref{figure_4} suggest that the situation is 
similar for the B supergiants, since at a given \Teff\, the estimates of 
\citet{fraser10}  for extra broadening (red dots) are systematically lower  
than those obtained by \citet{lefever07}, \citet{MP}, and \citet{simon10} 
(green dots). Indeed, from linear fits to the two datasets (dotted lines), 
we estimated a mean difference of 19$\pm$4~\kms, in perfect agreement with 
our estimate for the ESO O-star sample and results from \citet{dufton06a} 
for B supergiants in the SMC. Following these findings, we added a constant 
value of 19~\kms to the extra line-broadening velocities adopted from 
\citet{fraser10} and \citet{bouret12}, to make them consistent with the rest 
of the OB sample. 

Regarding the \Teff\, and \logg\, values of the stars included in the OB sample, 
the majority of them have been derived by applying the non-LTE, line-blanketed 
model atmosphere codes (allowing for spherical extension and stellar winds) 
FASTWIND (present study,  \citealt{lefever07, MP, markova08}) or CMFGEN \citep{bouret12, 
searil}. Only in \citet{fraser10}, the non-LTE, line-blanketed, plane-parallel 
model atmosphere and line formation codes TLUSTY and SYNSPEC (see, e.g. 
\citealt{H88, HL95, LH07}) were used instead.  While  the overall agreement 
between these codes (within their domains of application) was generally 
considered as satisfactory (see, e.g., \citealt{puls05, crowther06, MP}), in a 
recent study  \citet{massey13}  noted a systematic difference between O-star 
surface gravities derived by either FASTWIND or CMFGEN, with the former being 
smaller by about 0.1~dex. To account for these new results, a possible asymmetric 
error of +0.2/-0.1 dex in \logg\, was considered for all stars in the OB sample.   

\subsection{Projected rotational and extra line-broadening 
velocities as a function of effective temperatures} 
\label{teff_dependence} 

\begin{figure}
{\includegraphics[width=8.5cm,height=5.5cm]{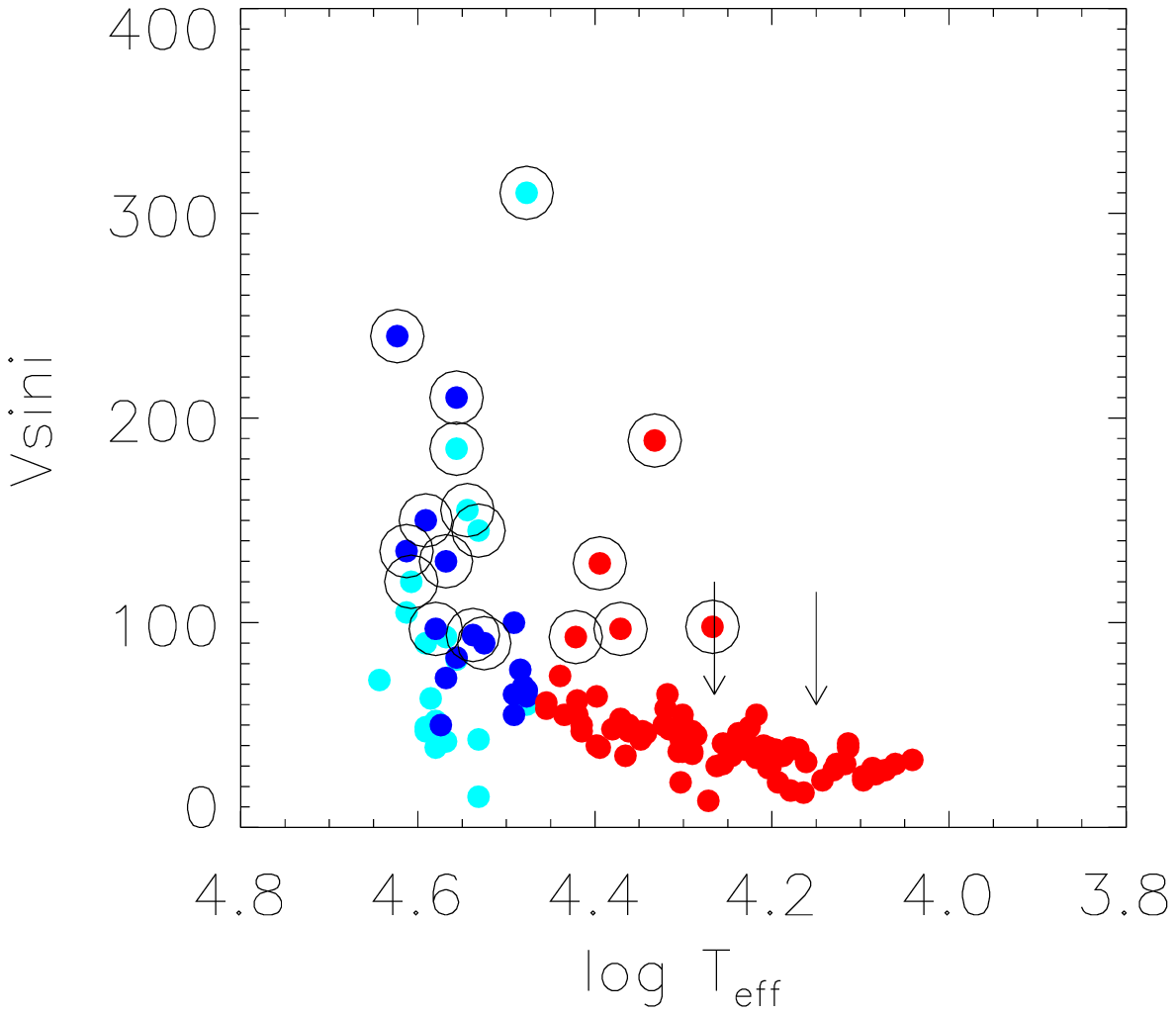}}
\\
{\includegraphics[width=8.5cm,height=5.5cm]{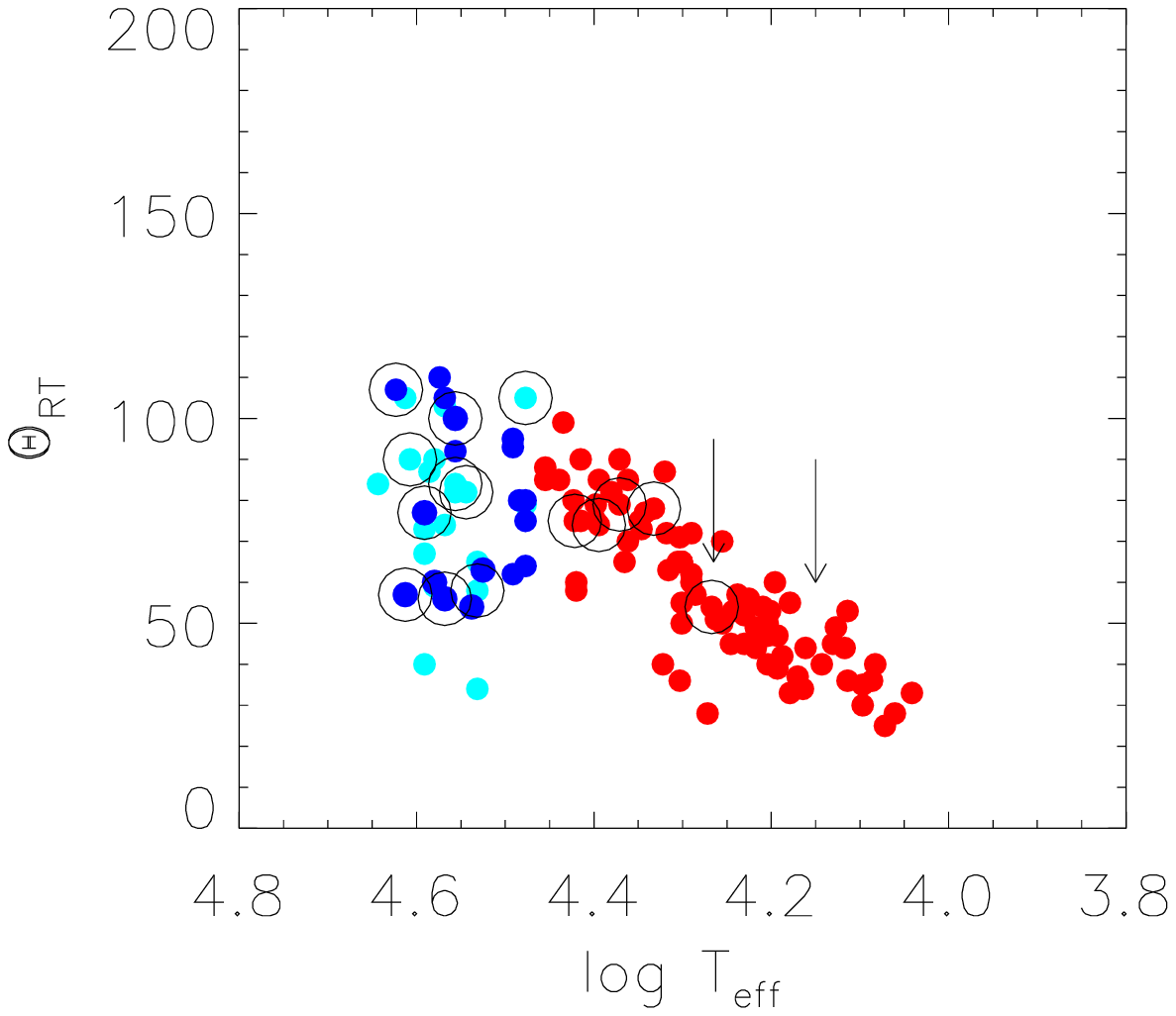}}
\caption{Projected rotational (upper panel) and extra line-broadening 
velocities (lower panel) for the OB sample as a function of \Teff.  O 
and B supergiants are highlighted in dark blue and red, respectively, 
and O giants and dwarfs in light blue. Intermediate- and fast-rotating 
stars (\vsini$\ge$110~\kms)  are additionally marked by large circles 
(for more information and arrows  see Sect.\ref{teff_dependence}).}
\label{figure_5}
\end{figure}

The upper panel of Fig.~\ref{figure_5} displays the run of the \vsini\, 
values for the OB sample as a function of \Teff.  On a global scale, 
a general trend of decreasing projected rotational velocities with 
decreasing temperatures is observed, where the hotter stars are 
spread over a wider range of values than the cooler ones. Since 
massive B supergiants are considered as evolutionary successors of O-type 
stars, this result may indicate that  -- in agreement with evolutionary 
predictions \citep{mm00, brott} -- the surface equatorial velocities 
of hot massive stars decrease continuously while crossing the 
Hertzsprung-Russell diagram (HRD), with a scatter being caused  by 
projection and differences in stellar masses and initial rotational rates.

Additionally, our data reveal that while the hotter stars appear 
to  demonstrate a deficit of very slow rotators - we have only one 
star with \vsini\, below 40~\kms - those at the cooler edge of 
the B-supergiant domain lack fast rotators\footnote{For the 
B-supergiant sample we define fast rotators  as those with  particularly 
high  \vsini\, for their corresponding \Teff.}, with a border line between 
the two subgroups located at log~\Teff\, of 4.25-4.30 dex. 

The steep drop in massive star rotational rates at \Teff\, of about 
22\,000~K has been highlighted by \citet{vink10} based on the dataset 
of \citet{howarth97} for \vsini, which  does not account for extra 
broadening (see their Fig.2). To explain this feature, the authors  
suggested two potential scenarios: enhanced mass loss at the predicted 
location of the bi-stability (BS) jump, or post-MS angular momentum loss.

Regarding the deficit of slowest projected rotators among the hotter 
stars in the OB sample, we mentioned in Sect.~\ref{vrot_oldnew} that
this feature might be due to selection effects (we recall that our \vsini\,
were derived accounting for extra broadening). Indeed, a comparison 
of Fig.~2 from \citet{vink10}  and the upper panel of our Fig.~\ref{figure_5} 
reveals that while in  the sample of \citet{howarth97}  there are a large 
number of slowly rotating, low-luminosity O stars, such stars are missing 
in our case. On the other hand, it is clear that this finding does not 
apply to supergiants: both samples demonstrate a deficit of narrow-lined 
objects toward higher \Teff. Thus, we may conclude that reasons different 
from observational selection must be present and contribute to the deficit 
of slowest projected rotators among the hotter and more luminous stars in 
the OB sample.  This matter is discussed in more detail in the next section.

In the lower panel of Fig.~\ref{figure_5}, we show the run of \eb\, with  \Teff.
Interestingly,  the decline of \eb\, toward cooler \Teff\, established for 
the supergiants in our basic  O-star sample  (see Fig.~\ref{figure_3}) becomes 
questionable with the inclusion of the targets of \citet{bouret12} at 
intermediate and fast rotation, leading to broadening velocities quite similar 
to those observed for the hotter B supergiants. While the explanation of this 
finding is currently unclear, one can see that the \eb\, velocities for the 
cooler B supergiants are definitely lower that those for the O stars and  hotter 
B supergiants, suggesting that changes in the evolutionary stage of hot 
massive stars may play a role in determining the properties of extra broadening. 

Comparing the upper and lower panels of Fig.~\ref{figure_5}, the run of \eb\, 
with \Teff\, is surprisingly similar to that of \vsini, even regarding details. 
This refers for instnce to the spread of O-type dwarfs and giants toward lower 
\vsini; the presence of two targets with peculiarly low \vsini\, for their \Teff\, 
(HD~159110, log~\Teff~=~4.27, \vsini = 9~\kms and HD~52089, log~\Teff = 4.3, 
\vsini =  22~\kms); and the two local depressions in the \vsini\, distribution, 
marked by vertical arrows. These similarities are quite interesting because they 
may indicate that the two broadening agents are linked (this question is  
addressed in more detail in Sect.~\ref{rot_extra}).
\begin{figure}
\begin{center}
{\includegraphics[width=9.5cm,height=6.5cm]{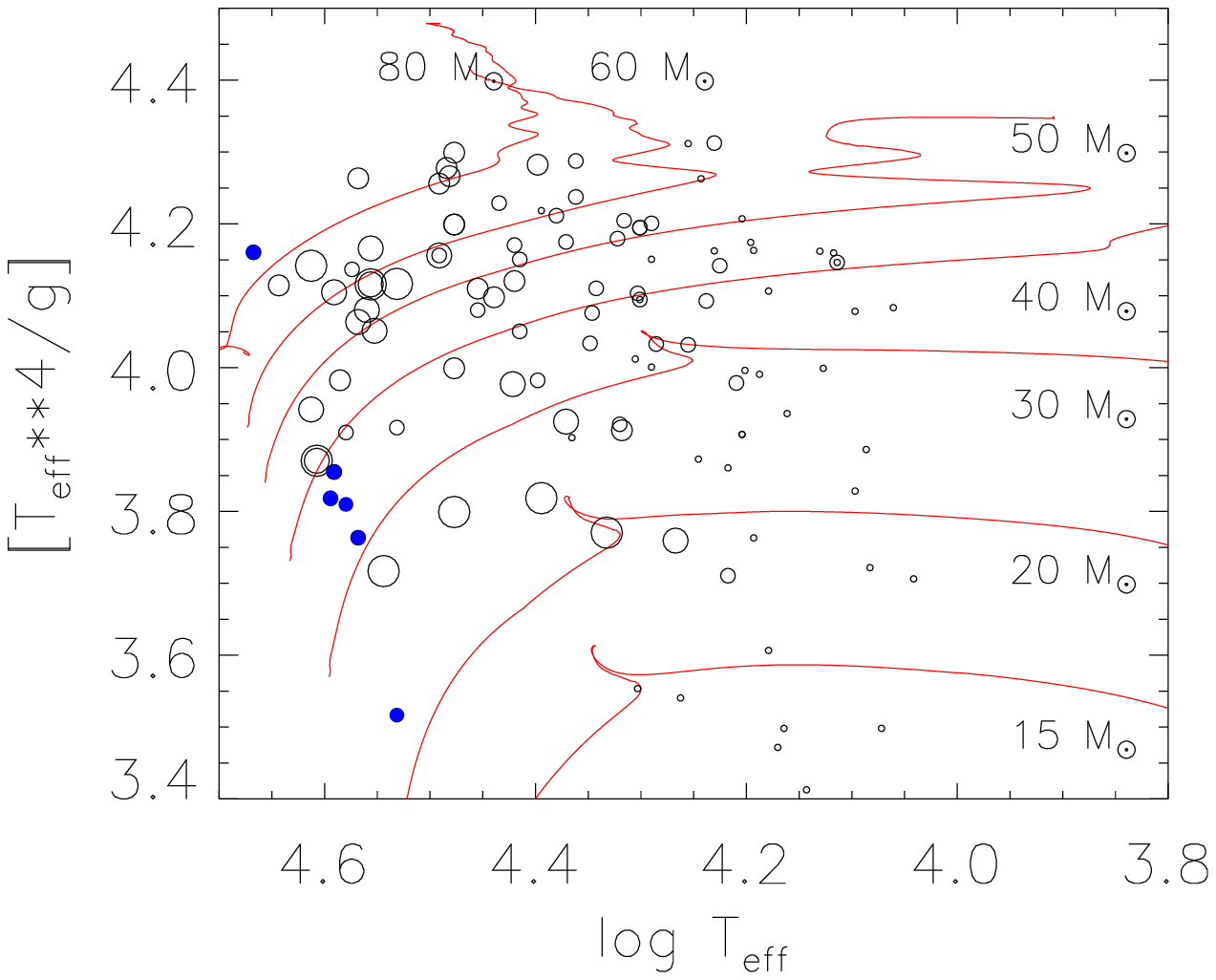}}
{\includegraphics[width=9.5cm,height=6.5cm]{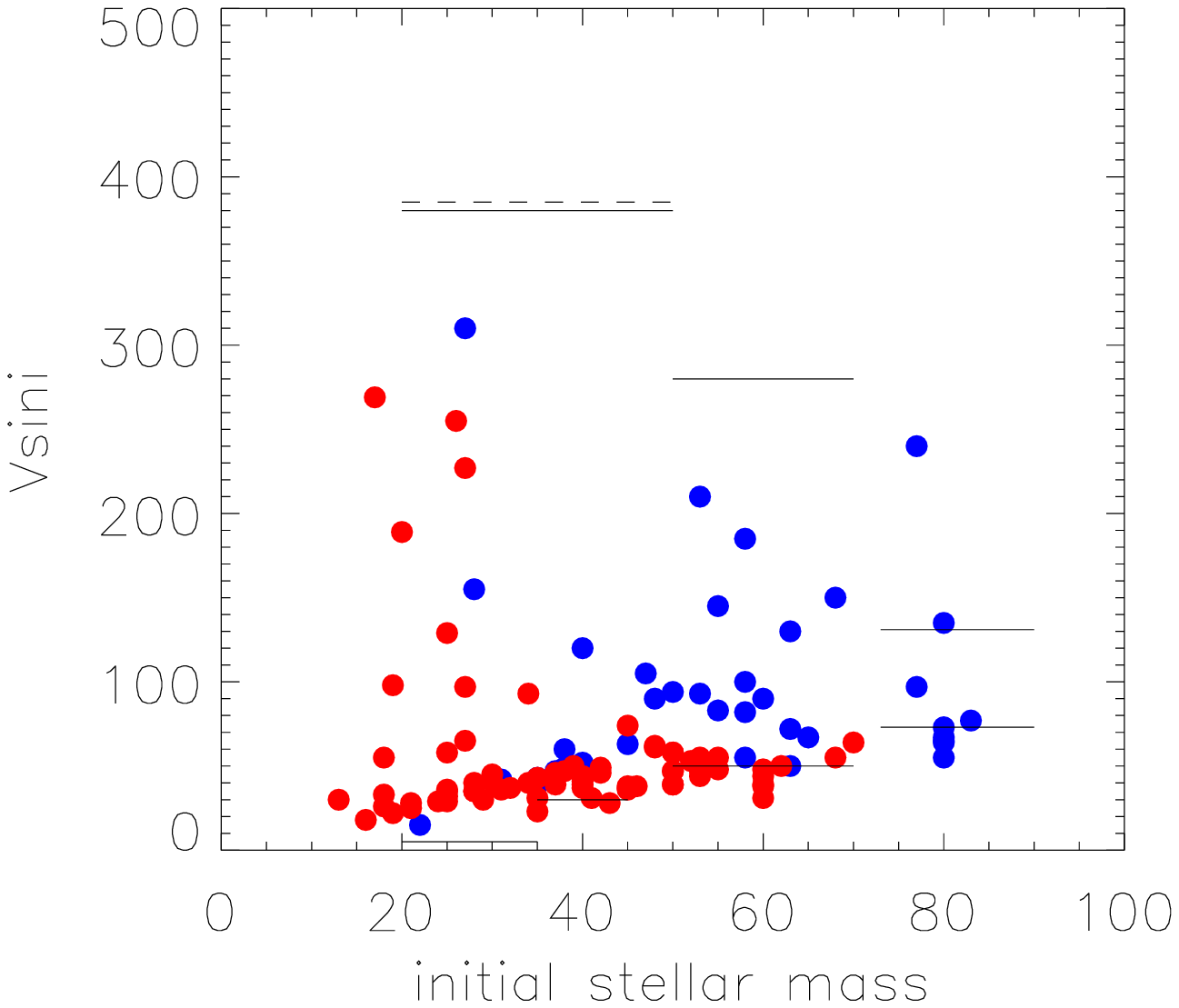}}
\caption{{\it Upper panel}:  Values of $\Teffe^{4}/g$ vs. \Teff for 
our OB sample, with
$ \left [ \Teffe^{4}/g  \right ]
= \log \left ( \Teffe^{4}/g \right ) - \log \left ( \Teffe^{4}/g \right )_\odot$.
Overplotted are the evolutionary tracks from \citet{brott} 
for \vinit$\approx$300~\kms and seven values of \Minit. The size 
of the data points is proportional to the values of \vsini. Stars close 
to the ZAMS with anomalously low \vsini\, are highlighted in blue. 
\newline {\it Lower panel}: Projected rotational velocities  vs.  initial 
evolutionary masses. O stars are colored in blue, and B-supergiants in red. 
Horizontal lines represent upper and lower limits to \vsini\, as 
deduced from Fig. 1 in \citet{wolff06} (dashed) and Fig. 5 of 
\citet{penny96} (solid), additionally corrected for by a constant value 
of -19~\kms to account for extra broadening.
}
\label{figure_6}
\end{center}
\end{figure}

\subsection{Projected rotational  velocities as a function of 
stellar mass and evolution}\label{vrot_evol}

\subsubsection{Dependence on stellar mass}
\label{obs_evol}

The upper panel of Fig.~\ref{figure_6} depicts the distribution of the 
OB sample in a $\log \Teffe^{4}/g$ versus $\log \Teffe$ diagram,  with 
symbol sizes proportional to the values of \vsini\footnote{As in a 
$\log g$ vs. $\log \Teffe$ diagram, the knowledge of the stellar 
distance is not required to  include stars in this diagram. Furthermore, 
it has the property that for stars with normal helium abundance, 
$\Teffe^{4}/g$ is proportional to their Eddington factor $\Gamma_e$, 
such that a value of $ \left [ \Teffe^{4}/g  \right ] \simeq 4.6 $ 
corresponds to $\Gamma_e =1$. Square  brackets denote logarithmic values in units
of solar ones.}. Overplotted are the evolutionary tracks from \citet{brott} 
for $Z$~=~0.02, with initial equatorial rotational velocities (\vinit) ranging 
from 312~\kms\, for the 60~\Msun\, model  to 329~\kms\, for the 
15~\Msun\, model, and an initial mass (\Minit) of 15, 20, 30, 40, 50, 60, 
and 80~\Msun. 

From these data we find  that all the stars in the OB sample lie in a region 
between $\sim$15 and $\sim$80~\Msun, with an obvious deficit of MS objects 
with \Minit$\lesssim$40~\Msun. Thus, our sample can be considered as 
representative only for the more massive OB stars. 

Additionally, we see that while objects close to the zero-age main sequence 
(ZAMS) can have quite different \vsini\, (most likely due to projection), 
there is, on a general scale, a progressive decline of \vsini\, toward later 
evolutionary stages. Suggestive evidence for the cooler, less massive B 
supergiants showing somewhat lower \vsini\, than the most massive ones at 
the same \Teff\, seems to be present as well. While the former result is 
qualitatively consistent with evolutionary predictions, the latter is somewhat 
confusing, because according to the models, the opposite effect is to be expected
(see \citealt{mm00, brott}).

To investigate this problem in more detail, in the lower panel of Fig.~\ref{figure_6} 
we display \vsini\, as a function of \Minit: O-type stars are highlighted 
in blue and B-supergiants in red. For each individual star, \Minit\, is 
determined by interpolating between the available evolutionary tracks. We 
are aware that \citet{fraser10} and \citet{bouret12} provided their own estimates 
of \Minit\, for their sample stars. However, since these data were obtained 
using  evolutionary tracks from \citet{mm03},  we performed our own determinations 
of \Minit\, for these objects as well for consistency reasons\footnote{Since in the 
Bonn models helium is practically not mixed out (unless in the limit of 
homogeneous evolution), while in the Geneva models it is mixed out continuously 
as a function of rotation, the former  become less luminous than the
latter, leading to somewhat higher \Minit\, at the same \Teff\, and \logg.}. For the 
stars listed in Table~\ref{sample}, \Minit\, are given in Column~8. The uncertainty 
in these data  reflects the errors on the effective temperatures 
and surface gravities and is on the order of -37/+25\% and -20/+12\%  for the 
higher- and  the lower-mass end, respectively\footnote{Interestingly, 
underestimated gravities lead to overestimated initial masses when derived 
via the $\Teffe^{4}/g$ vs. $\Teffe$ diagram, which basically reflects the 
impact of the mass-luminosity relation, with  an additional uncertainty 
being related to the accuracy of the model tracks and differences between 
the actual \vinit\, of the targets and the model value of about 300~\kms.}. 

One very conspicuous feature of  the distribution, shown  in the lower panel 
of Fig.~\ref{figure_6}, is the well-defined trend of increasing minimum 
\vsini\, toward higher masses. Indeed, for the O stars this feature is not 
clearly  manifested due to the limited number of objects with masses below 
50~\Msun. However, since a similar trend has been observed by \citet{penny96} 
for a statistically significant sample of O-type stars uniformly distributed 
over a mass range from $\sim$10 to $\sim$90~\Msun\, and by \citet{wolff06} 
for a comparatively limited (55 vs. 200 objects from Penny) sample of O stars 
with masses between $\sim$20 and $\sim$40~\Msun, and since for a given mass 
range their estimates agree quantitatively well with ours (see the horizontal 
lines shown in the lower panel of Fig.~\ref{figure_6}), we consider the 
following conclusion as reasonable: the distribution of \vsini\, as a function 
of \Minit\, for the OB sample is characterized by a lowest value of \vsini\, 
that increases toward higher masses, leading to a remarkable deficit of slowest 
projected rotators among the most massive objects.

Because  for a given initial mass hot massive stars are predicted to slow down 
when evolving to cooler temperatures, it follows from the lower panel of 
Fig.~\ref{figure_6} that high-mass stars do not evolve to velocities as low as 
those reached by low-mass ones. Using the  evolutionary calculations of 
\citet{brott} for \vinit$\sim$300~\kms, we investigated the evolutionary status 
of the stars tracing out the trend, and found that while  those with \Minit\, 
between 60 and $\sim$80\Msun\, appear to be at the end of the MS phase with ages 
between 2.8~-~3.6 Myr, those  with \Minit\, between 15 and less than 40\Msun\, 
might be considered as post-MS objects at the end of the B-supergiant phase, 
with ages ranging from $\sim$ 4.5 to $\sim$12~Myr (provided they all are moving 
to the red supergiant phase, but see below). Since the surface equatorial 
velocities of less-massive stars after the MS are predicted to drop very fast 
due to rapid expansion after core hydrogen exhaustion \citep{mm03,brott}, one 
may speculate that differences in evolutionary status can provide a clue for 
understanding the connection between the minimum \vsini\, and  \Minit\, for 
the stars  from the OB sample. 

There are, at least, two problems with this hypothesis: $first$, the models do 
not predict the presence of stars with $\log T_{\rm eff} \leq 4.25$ because
of the short lifetimes in these phases (see next section), and $second$, 
differences in evolutionary stages cannot account for the $complete$ lack of 
slowest projected rotators among the most massive stars because such objects 
are expected to be present in our sample as well as in the studies by 
\citet{CE77}, \citet{penny96}, \citet{howarth97}, and \citet{wolff06}, due to 
projection effects and stars with intrinsically low \vinit.

One way to  address this puzzling problem is to challenge the conventional 
view of randomly oriented rotational axes, assuming that those of the most 
massive OB supergiants are preferentially aligned with the Galactic poles. 
While this hypothesis may seem speculative and difficult to argue for, we 
mention that a similar possibility has been discussed in previous investigations 
regarding unusual \vsini\, distributions observed for supergiants of late 
spectral types (see \citealt{gray87} and references therein).  

From a different point of view, one  may assume  that the FT method fails to 
detect low values of \vsini\, due to an effect that is notable for the 
more massive objects and negligible for the less massive ones. Since 
instrumental effects cannot be responsible -- note that all data included in 
our analysis originate from high-quality observations allowing reliable 
estimates to be obtained down to \vsini\, of 10 to 15~\kms -- and since extra 
broadening has been taken into  account, we suggest that by introducing 
additional zeros in the FT, microturbulence might explain  the lack  of slow 
rotators among the more massive sample stars (\citealt{gray73}; see also 
\citealt{simon07}). The  problem with this explanation, at least at present, 
is that while it requires \vmic\, to increase toward higher masses for the 
OB sample, no evidence of a mass-dependent microturbulence was found when 
analyzing the corresponding data. On the other hand, it is worthwhile to note 
that microturbulent motions that become stronger toward higher masses and lower 
temperatures could result from subsurface convection caused by the iron opacity 
peak \citep{cantiello}. Clearly, additional effort is required to understand 
these observational results and to obtain a coherent picture about the nature 
and the origin of the deficit of slowest rotators among more massive OB stars
\footnote{We refer to \citet{SH13} for an in-depth discussion on the possible 
explanation of the lack of low \vsini\, stars due to the effects of 
microturbulent broadening in the case of OB-type stars.}.

Another intriguing feature regarding the \vsini\, properties of hot 
massive stars is that in our sample as well as in the samples studied
by previous investigators (e.g., \citealt{CE77, penny96, howarth97, 
wolff06}) there are hardly any O~stars with \vsini\, in excess of
400~\kms. For the particular case of stars with masses between 20 and
40~\Msun, located close to the ZAMS, \citet{wolff06} argued that the
400~\kms limit may be representative for their initial velocities.
How this argumentation applies to O stars with masses above 40~\Msun\,
is currently unknown. However, from the data shown in the lower panel
of Fig.~\ref{figure_6} and Fig.~5 in \citet{penny96}, it appears that
these stars tend to rotate generally more slowly than the less massive
ones, with rapid rotators being extremely rare amongst them (see the
horizontal lines shown in the lower panel of Fig.~\ref{figure_6}). We
return to this matter in the next section.

\subsubsection{Comparison with evolutionary predictions}
\label{comp_mod}

\begin{figure}
{\includegraphics[width=8.5cm,height=5.5cm]{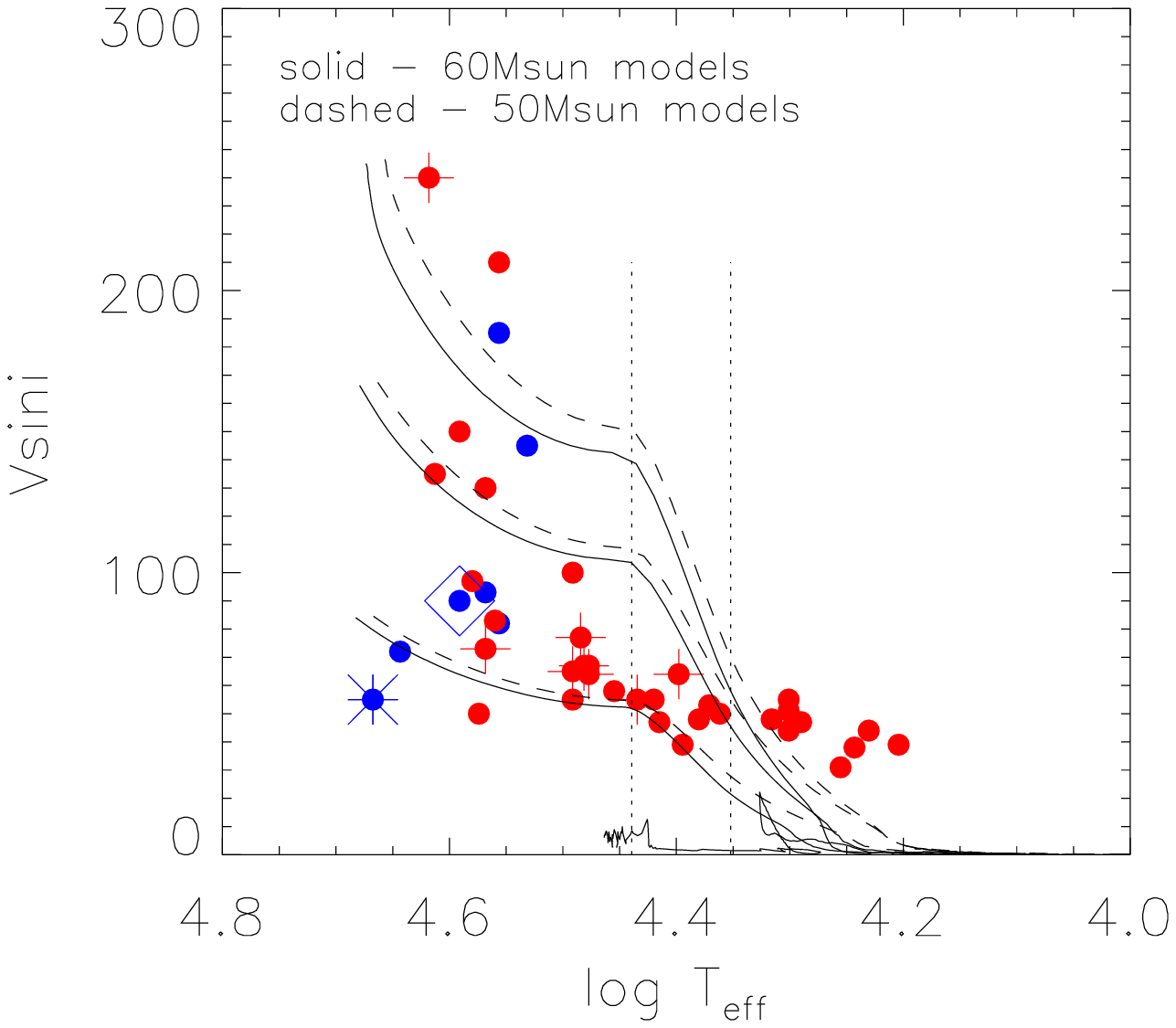}}
{\includegraphics[width=8.5cm,height=5.5cm]{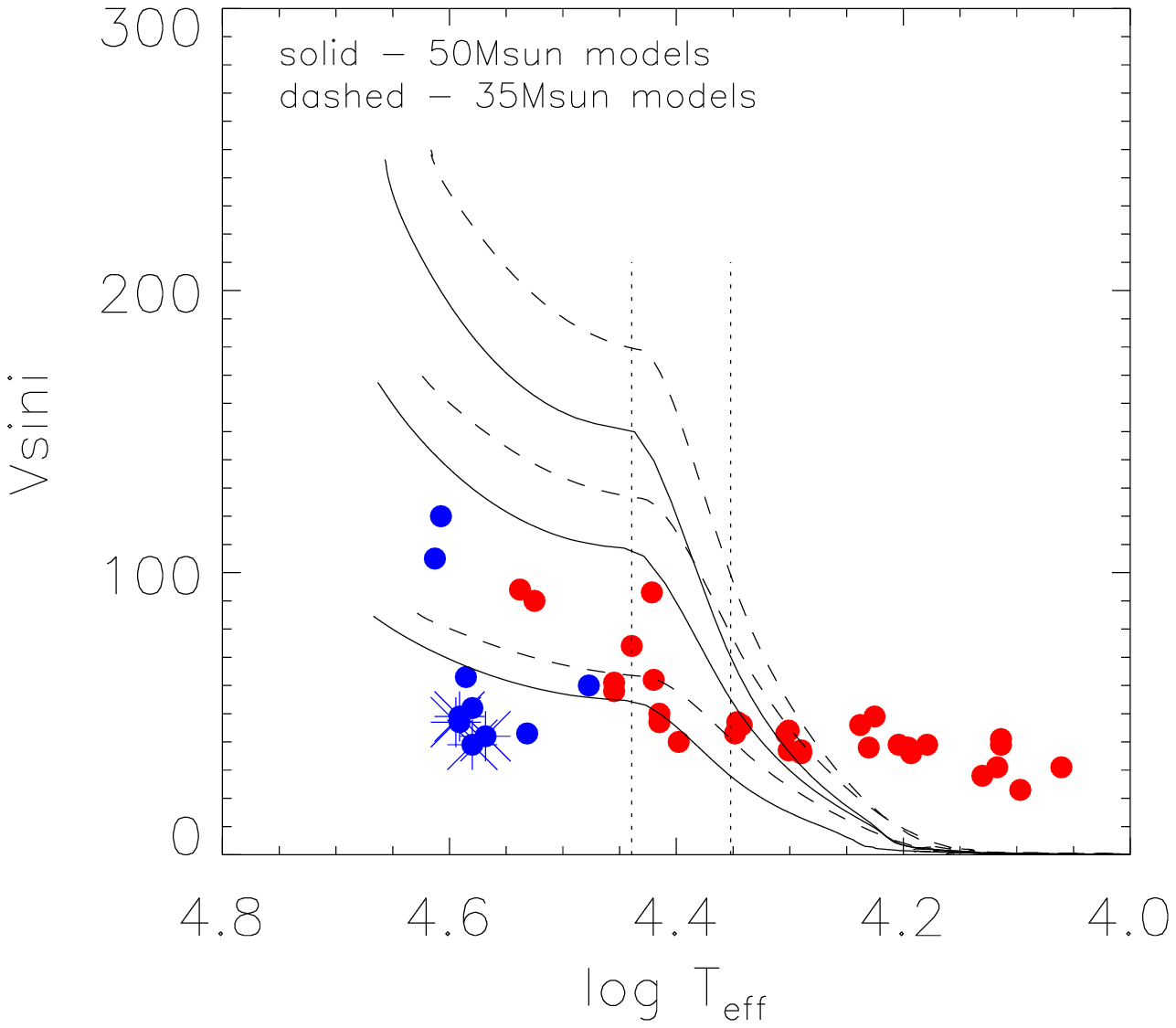}}
{\includegraphics[width=8.5cm,height=5.5cm]{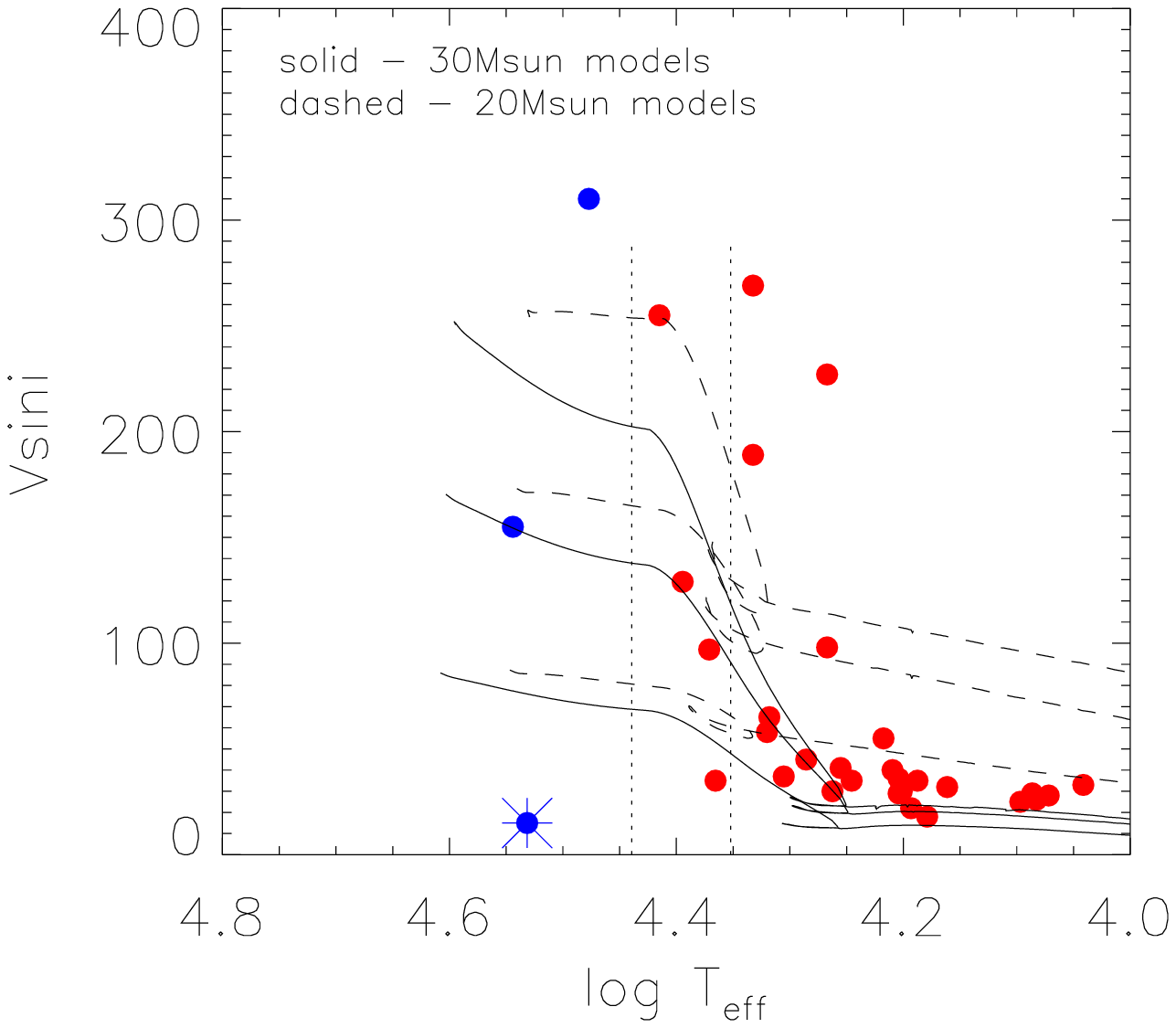}}
\caption{Projected rotation velocities and effective temperatures 
for the OB supergiants (red) and the O-type giants and dwarfs (blue)
from our sample, plotted against evolutionary tracks 
from \citet{brott}. All stars have been binned into three mass 
ranges: 15\Msun$\le$M$\le$30\Msun (bottom panel); 
31\Msun$\le$M$\le$50\Msun (middle panel), and M$\ge$51\Msun (top panel). 
O-type stars close to the ZAMS with anomalously low \vsini\, are 
additionally highlighted by asterisks. The large diamond marks 
HD~93843, which has been recently identified as a magnetic star 
\citep{hubrig11}. In the top panel, the most massive stars 
(\Minit$\gtrsim$70\Msun) are marked by crosses. The model values have 
been scaled by $\pi$/4 to account for projection. The two vertical 
lines represent the temperature limits of the bi-stability jump, as 
determined by \citet{vink99}. 
}
%
\label{figure_7}
\end{figure}

In Fig.~\ref{figure_7} we show the \vsini\, vs. log \Teff\, distribution 
for the OB sample against the model tracks from \citet{brott} for five values 
of \Minit (quoted in the upper left corner of each plot), and three values of 
\vinit: $\sim$100, $\sim$ 200 and $\sim$300~\kms. All tracks were scaled by 
a factor $\pi$/4 to account for projection, assuming a random orientation of
rotational axes. To facilitate the comparison, all stars were binned into three 
mass ranges: from 15 to 30~\Msun\, (bottom panel), from 31 to 50~\Msun\, (middle 
panel), and from 51 to $\sim$80~\Msun\, (top panel). Since for \Minit$\ge$ 60~\Msun\, 
no model predictions for the evolution of rotation are available,  the sample
stars with masses above 70~\Msun are highlighted with a different symbol in the 
top panel. The dotted vertical lines mark the position of the bi-stability (BS) 
region, as predicted by \citet{vink99} (but see Sect.~\ref{vmac_evol}).

Before outlining our results, we note that particularly the models of \citet{brott} 
are interesting to compare with ours, since the overshooting parameter used in 
these models was calibrated such that the $\log g$ value at the terminal age 
main-sequence of 10...20~\Msun stars corresponds to the observed $\log g$ value 
at which the LMC B~stars' rotational velocity distribution as a function of  
$\log g$ shows a sudden drop in the mentioned mass range. While this leads to a 
wider main sequence band than was, for instance  found in the Geneva models 
(\citet{mm03}), the evolution of the surface rotational velocity with time 
is qualitatively similar in both types of models.

\paragraph{\it The case of more massive stars}
From the data illustrated in the top and middle panel of Fig.~\ref{figure_7}, 
we see that the models with \Minit$\ge$35\Msun\, predict a significant 
decline in \vsini\, before the BS-region, followed by  a second steep 
decline within this region,  and values going to zero after that (the 
so-called BS braking, see \citealt{vink10}). By confronting model predictions 
with the observations, we furthermore find that
\begin{itemize}
\item[i)] $before$ the BS-region (\Teff$\ge$4.44~dex), the  agreement between 
theory and observations is quite satisfactory, with the majority of datapoints  
being located between tracks with \vinit\, of 100 and 200~\kms,  and individual 
cases clustering around the 300~\kms model. Since all these stars appear to be 
relatively young MS objects with ages ranging between one and 4.0 Myr, and since 
their number seems to be too large 
to be solely explained by projection effects\footnote{Under the assumption of 
random orientation of rotational axes,  only 30 percent of the stars will be 
located in the first half of the distribution, at $i$ between zero and 45~deg.}, 
we suggest that a large part of them may have been rotating much slower than 
critical when appearing at or close to the ZAMS  rather than been spun down 
within the first few Myr of evolution. 
\item[ii)]  $after$ the BS-region (\Teff$\le$4.35~dex), a significant 
discrepancy between the measured and predicted rotational rates is observed,  
where the former do not go to zero, as predicted  by the models, but display 
a mild decline toward lower \Teff, being at the same time systematically 
higher than the model values. This result might  imply that for more massive 
stars the ejection of angular momentum during the BS-jump has been overpredicted 
(due to an overpredicted jump in \Mdot, for example see \citealt{MP}), leading to 
significantly lower rates than observed. Alternatively, and related to our 
discussion in Sect.~\ref{obs_evol}, the FT method may have led to an overestimate
of \vsini\, for slow rotators, in particular for the cooler and more 
massive B supergiants.
\end{itemize}

Comparing the observed values of \vsini\, with critical velocities, 
\vcrit, as calculated by \citet{brott}, we find that the most massive 
stars (\Minit$\gtrsim$50~\Msun) in the OB sample may appear at or close to 
the ZAMS with velocities that do not exceed 26\% of their \vcrit; 
for the majority of them  this limit might be even lower, about 17\%.  
Similar findings  for stars with masses between 10 and 40~\Msun\, indicate  
an upper limit of $\sim$50\%,  with most of the stars rotating at less 
than 30\% of the critical rate \citep{wolff06}. Taken together, these results  
suggest (i) that the initial velocity of hot massive stars might  
be mass-dependent, and  (ii) that some unique 
mechanism must be at work to keep the rotational rates of these stars low 
at birth. A potential candidate for this role is gravitational torque, 
which seems to be able to solely prevent the central object from spinning-up 
to more than half of its break-up velocity \citep{lin11}. For high-mass 
stars, intensive mass loss during the first one Myr - when the stars might 
be obscured by gas in their parental clouds and thus are inaccessible for 
observation - might additionally contribute to spin these objects down to 
even lower velocities.

During the past decade two mechanisms have been proposed  to explain the 
formation of massive stars: via mergers of low-mass protostellar cores  
\citep{BZ05} or via a single lower-mass core that grows in mass via 
competitive accretion of surrounding molecular gas \citep{bonnell}. Since 
in the former model the result of the mergers is predicted to rotate rapidly 
\citep{BZ05}, while in the latter the final product is expected to rotate 
relatively slowly, at a speed well below the critical velocity \citep{shu94, 
lin11}, the very low rotational velocities reported here and in \citet{wolff06} 
suggest that a single core formation and not a core merger is the likely 
mechanism for massive stars to form. Studying the rotational rates of a large 
sample of single O stars in the LMC, \citet{ramirez} came to the same conclusion. 

Because significant effects for the evolution of massive single stars, including 
their potential to develop a long-duration gamma-ray burst at the end of 
their life, are expected for \vinit\, above 200$\dots$300\,\kms \citep{brott, 
langer12}, these results may also suggest that such effects are weaker then 
previously thought --- unless a massive star is spun up later in its life due 
to close binary interaction.

\paragraph{\it The case of less massive stars}
From the bottom panel of Fig.~\ref{figure_7}, one may conclude that while 
on the hotter side of the BS region no constraints on model predictions can 
be put due to the lack of less-massive hot stars in the sample, on the cooler 
side  and $inside$  this region the observed \vsini\, agree generally well 
with the predictions - with the majority of datapoints clustering between the 
tracks with \vinit\, of $\sim$100 and $\sim$200~\kms. 

However, we note that while Fig.~\ref{figure_7} does show how the rotational 
velocity of the models evolves after the end of core hydrogen burning, a 
comparison of the corresponding part of the tracks with the observed stars at 
$\log T_{\rm eff}\leq4.25$ is not meaningful, because the evolutionary timescale 
of the models is very short, and practically no stars are predicted to exist 
at these temperatures.  Note also that this is not a particular feature of the
models of \citet{brott}.  \citet{fraser10}, for instance,  compared their \vsini\,
determinations for Galactic B~supergiants with models from \citet{mm03}. These 
models also expand rapidly after core-hydrogen exhaustion, and a comparison of 
their post-main sequence rotational velocities with the observed values -- even 
though performed by \citet{fraser10} -- is basically meaningless. The 
interpretation of the rotation rates of B~supergiants in the considered mass 
range has to await stellar evolution models that predict their existence with a
frequency similar to that of evolved stars during core-hydrogen burning 
(cf. \citet{vink10}).

We conclude this section with comments on the slowly rotating O-stars in the 
OB sample relatively close to the ZAMS (upper panel of Fig.~\ref{figure_6}, 
datapoints colored in blue). These are: HD~64568, HD~91572, HD~91824, 
CPD\,$-$58\,2620, HD~97848, and HD~46202. Traditionally, these stars can be 
explained by projection effects.  However, there are recent evolutionary 
calculations \citep{asif09, m11}, supported by observational results \citep{t10}, 
which show that magnetic fields can effectively reduce the rotation rates of 
hot massive stars with line-driven winds (the so-called magnetic braking). 
With this possibility in mind, we searched the literature and found  that one 
of the noted six targets (CPD\,$-$58\,2620) has been investigated for the 
presence of magnetic fields, with no detection above the one sigma level, 
however \citep{hubrig11}. For another star, HD~93843, a mean longitudinal 
magnetic field above the 3$\sigma$ level has been detected  (but see also 
\citet{bagnulo12}), which might possibly explain its relatively low 
\vsini\, (cf. the large diamond in the upper panel of Fig.~\ref{figure_7}). 
On the other hand, following \citet{m11}, we find that if our six O-type 
dwarfs with peculiarly low \vsini\, had masses of 10~\Msun, then a magnetic-
field strength of one kG for solid body rotation and of one to three kG for 
differential rotation would be enough to spin them down to the observed \vsini. 
How these numbers would translate into the case of stars with masses above 40~\Msun\, 
is currently unclear. Thus, and at least at present, it does not seem possible 
to judge whether or not magnetic braking is responsible for the low rotational 
rates of our O-type dwarfs relatively close to the ZAMS.

\subsubsection{Some insights into the nature of low rotational rates 
of cooler B supergiants}\label{coolB}

\citet{vink10} highlighted the absence of rapid rotators among 
B-supergiants with \Teff\, below 22~kK in our Galaxy and in the LMC, 
discussing two possible explanations: wind-induced braking due to bi-stability, 
and post-MS angular momentum loss. While within the first scenario the stars 
on both sides of 22~kK are required to undergo core-hydrogen burning,  within 
the second one the cool B-supergiants would form a population entirely 
separated from the hotter MS stars.

To test these possibilities, we investigated the evolutionary status of our 
sample using evolutionary tracks from \citet{brott} with \vinit$\sim$300~\kms 
and found that all stars with \Minit$\ge$40~\Msun\, and $\log$\Teff$\gtrsim$~4.20~dex 
appear to be MS objects. These results suggest that changes of the wind properties 
in the BS region  can in principle be responsible for the deficit of rapid rotators 
among the cooler and more massive B-supergiants in our sample. Indeed, the fact 
that only one of the more massive stars located $inside$ the BS-region (see top and 
middle panels of Fig.\ref{figure_7}) has a \vsini\, significantly higher than those 
observed $after$ this region might lead to the conclusion that our data  do not 
give clear evidence of a steep drop in \vsini\, inside the BS-region. This conclusion, 
however, would be somewhat premature: reasons such as the established deficit of 
very fast rotators among more massive O-stars as well as the short time-interval 
needed for a more massive hot star to cross the BS-region (less than 
$\sim$4x10$^{5}$ yr compared to $\sim$2-2.5x10$^{6}$ yr for a star with a mass 
below 30~\Msun, \citealt{brott}) might equally contribute,  leading to the 
picture described above. Additionally, it is clear that at least for the stars 
with \vinit$\ge$200~\kms some kind of additional braking is necessary to spin 
these objects down to \vsini$\sim$60 to 50~\kms\, at $\log$\Teff$\sim$4.35~dex.
We note that the rotating models of \citet{mm03} in the mass range 30...50 end 
their core-hydrogen burning at $\log$\Teff$<$4.35 and thus predict a very small 
number of B~supergiants with lower effective temperatures.
\begin{figure}
{\includegraphics[width=8.5cm,height=5.5cm]{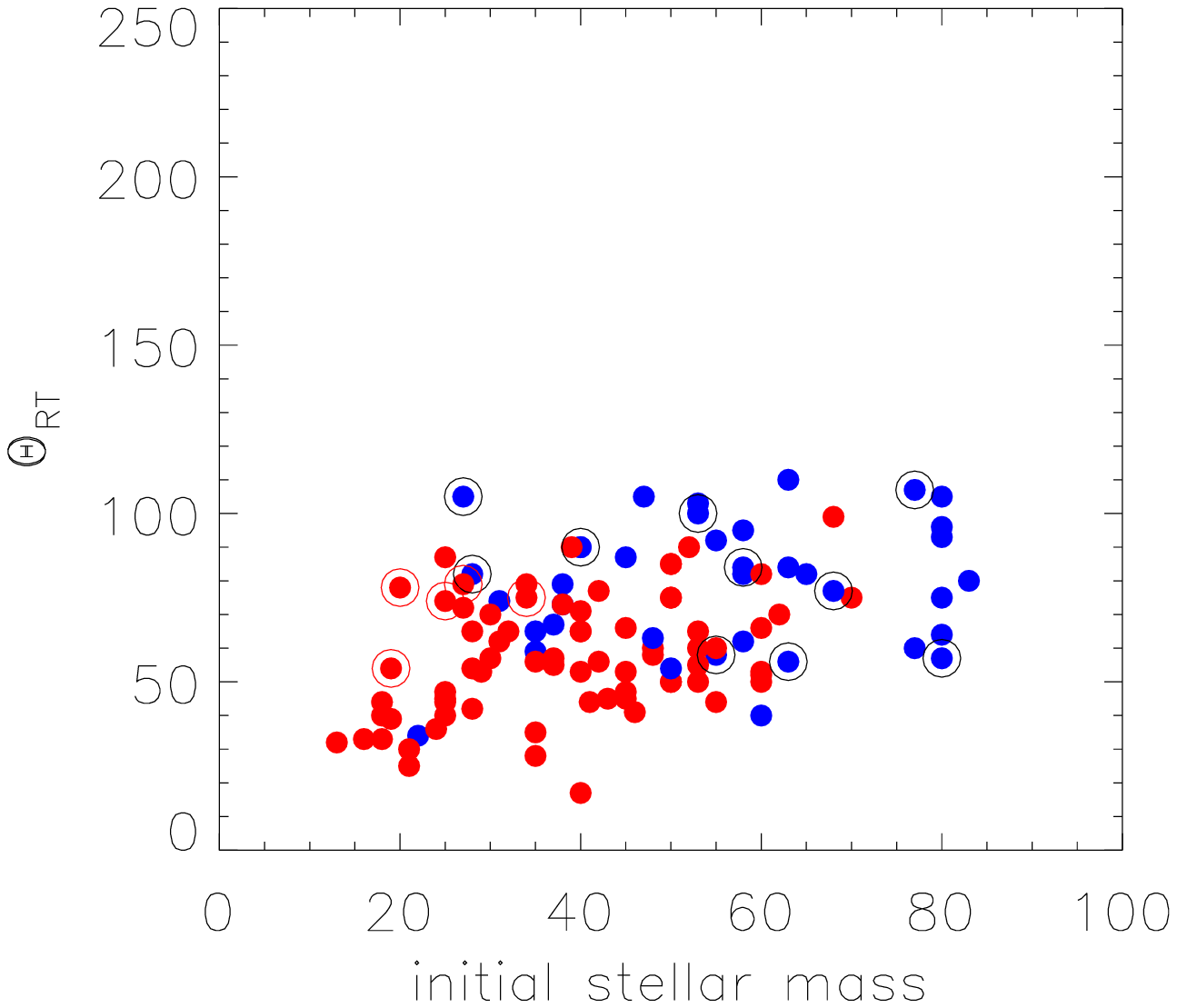}}
\caption{Same as the lower panel of Fig. 6, but for the extra line-broadening 
velocities. Intermediate and fast rotators are highlighted by large circles.
}
\label{figure_8}
\end{figure}
\begin{figure}
{\includegraphics[width=8.5cm,height=5.cm]{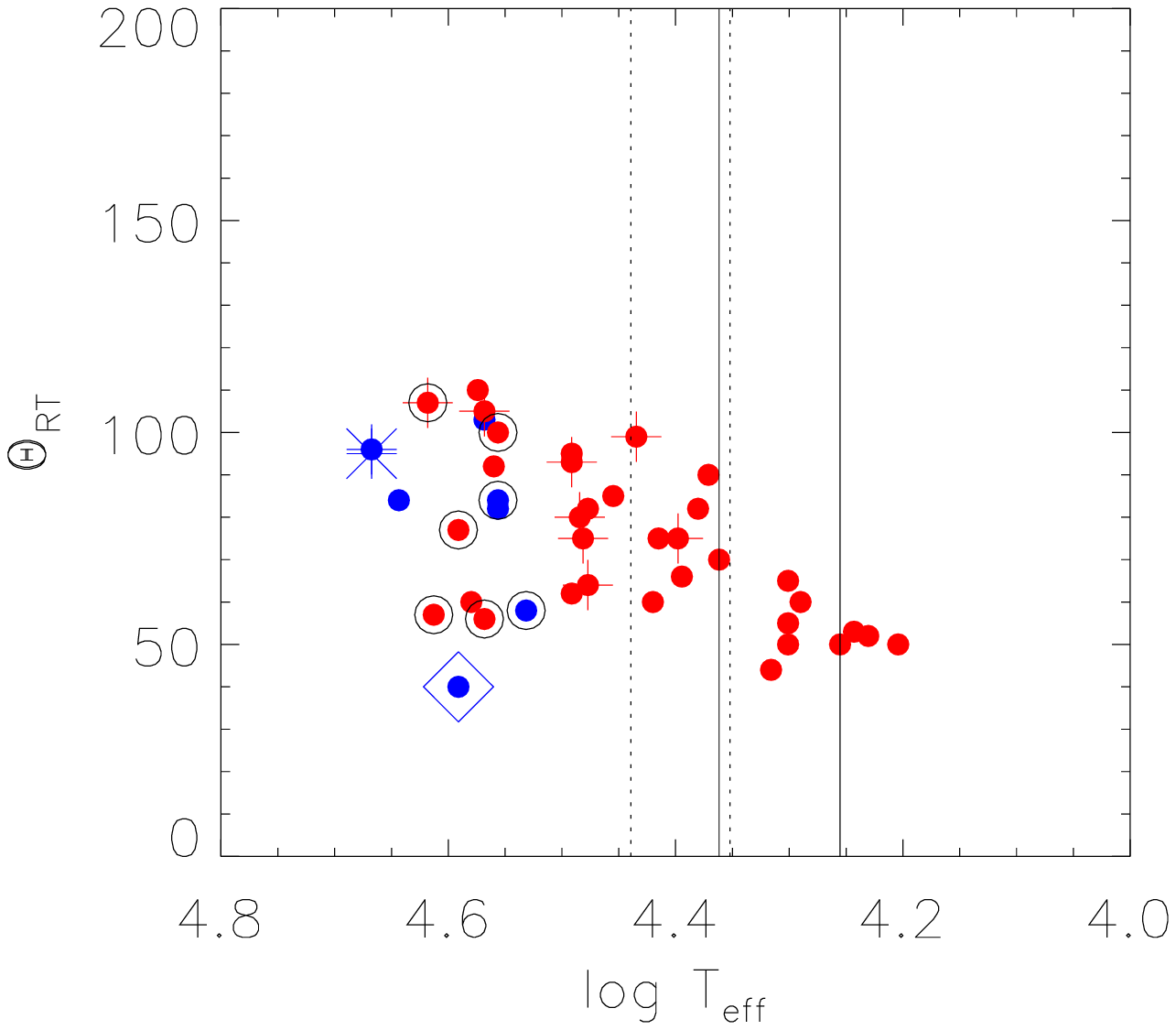}}
{\includegraphics[width=8.5cm,height=5.cm]{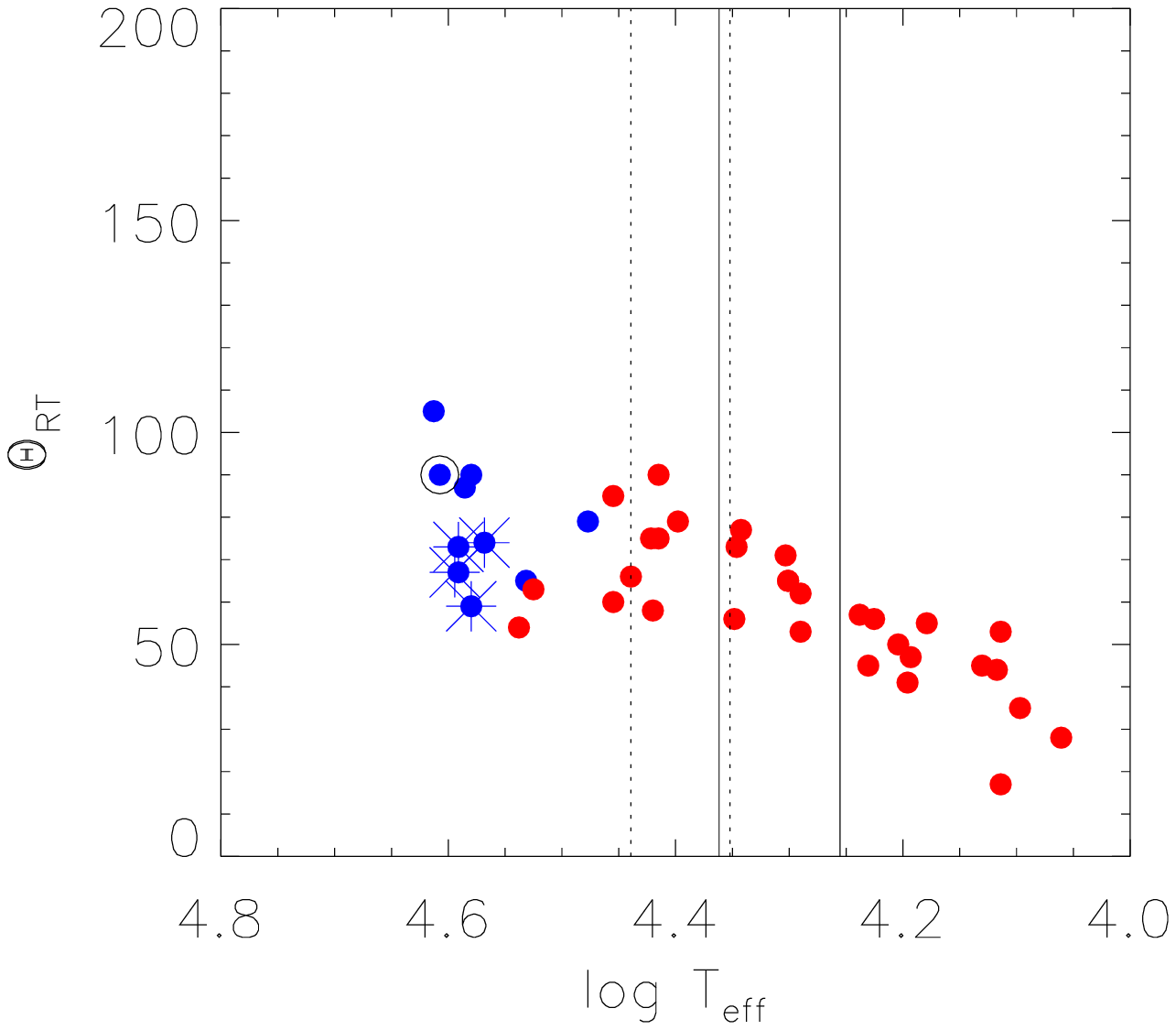}}
{\includegraphics[width=8.5cm,height=5.cm]{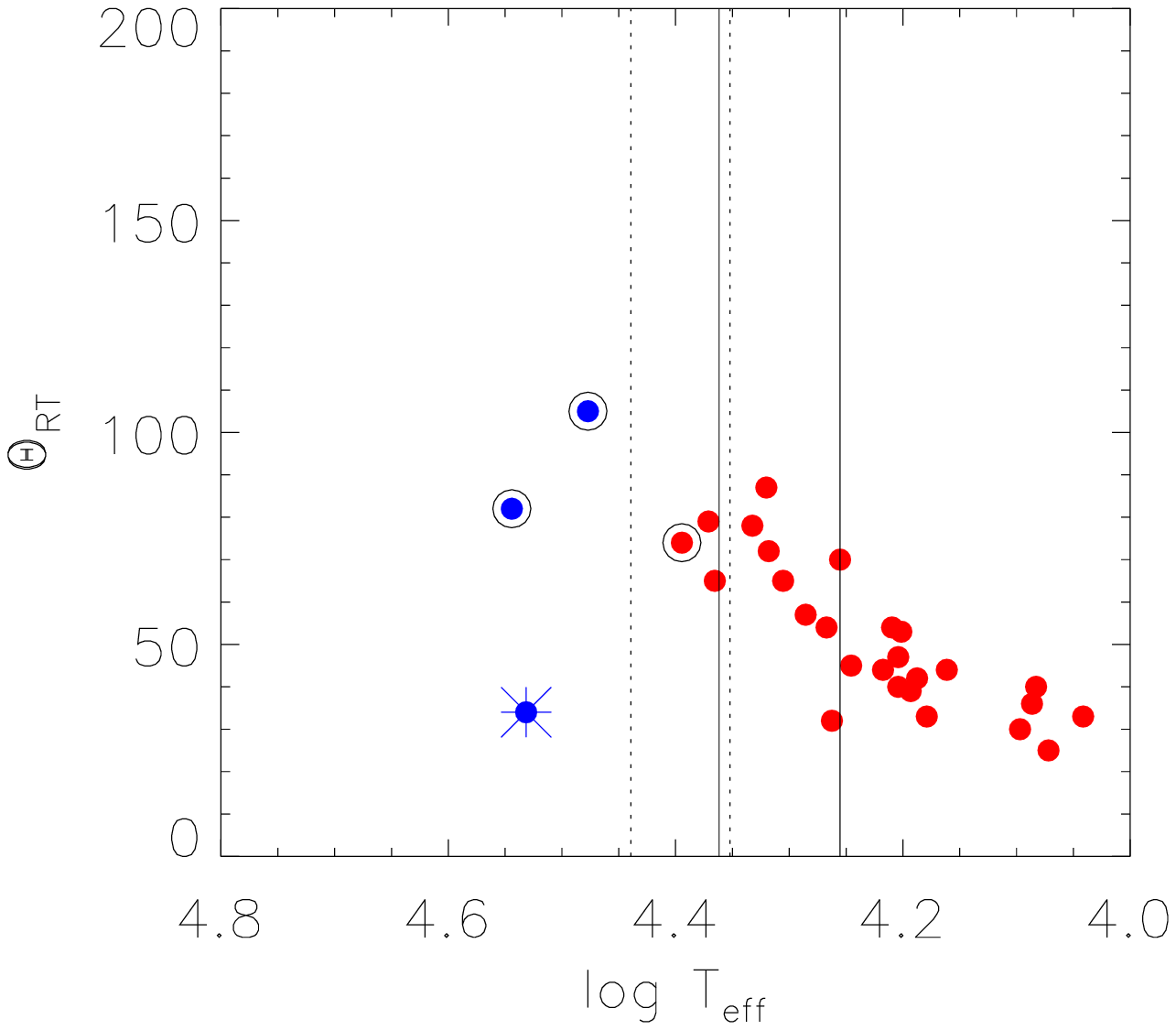}}
\caption{Extra line-broadening velocities as a function of \Teff\, for the 
stars from the three mass ranges as defined in Sect.~\ref{comp_mod}. 
Intermediate- and fast-rotating stars are additionally marked with 
large circles. The solid and dotted vertical lines mark the 
position of the observed and the theoretically predicted BS region, 
respectively. All other symbols have the same meaning as in
Fig.~\ref{figure_7}.}
\label{figure_9}
\end{figure}

\subsection{Extra broadening as a function 
of stellar mass and evolution}\label{vmac_evol}

Fig.\ref{figure_8} shows the \eb\, values for the stars from OB sample as a 
function of \Minit: again, O-type stars are highlighted in blue, and 
B-supergiants in red. Taken at face value, these data suggest that on a general 
scale the extra broadening rates of massive stars may decrease toward lower masses.  
Interestingly, for the intermediate and fast rotators no indication of a mass-dependent 
behaviour of \eb\, can be seen.  Due to the limited number of such objects in the 
OB sample, this result must be considered with caution, however.

To investigate the {\it evolution} of extra broadening when accounting for its 
possible dependence on stellar mass, we display  in Fig.~\ref{figure_9} \eb\, as 
a function of \Teff\, for the stars from the three mass ranges as defined in the 
previous section. From the top and middle panels of this figure, one may note 
that while for log\Teff~$\ge$ 4.35 the available data do not give evidence of a 
systematic change when the stars evolve to cooler temperatures, a progressive
decline toward later evolutionary stages is observed for the stars on the cooler 
side of this temperature limit.  Whether these findings apply also to the 
less-massive stars with \Minit $\le$ 35~\Msun\, (bottom panel of Fig.~\ref{figure_9}) 
cannot be judged at the momement due to the limited number of objects with 
log\Teff~$\ge$~4.35 and the problem outlined for the cooler B-supergiants in 
this mass regime (see Sect.~\ref{comp_mod}). 

Based on theoretical considerations, \citet{vink00} argued that the mass-loss 
rates of hot massive stars should increase steeply between log\Teff~$\approx$~4.35{\ldots}4.45 
because of the bi-stability mechanism. Since the predicted jump in mass loss is 
located at temperatures that are somewhat hotter than the one at which the extra 
broadening velocities change from higher and relatively constant values to values 
decreasing toward cooler temperatures (see the two dashed vertical lines in 
Fig.~\ref{figure_9}), one might conclude that changes in the wind properties when 
crossing the BS region cannot be responsible for or contribute to the observed 
behavior of \eb\, with \Teff. However, we recall that there is observational
evidence that indicates that the mass-loss rates of hot massive stars may not 
increase at the BS jump, and that the jump is located at somewhat cooler temperatures 
than those predicted by Vink et al. \citep{MP}. In this situation, the results 
shown in the top and middle panel of Fig.~\ref{figure_9} indicate that the BS jump 
plays a role in determining the properties of extra broadening in massive stars.
More investigations are required before one can judge which of the two
possibilities is the more likely one.

Finally, we return to our suggestion that the slow rotators relatively close to 
the ZAMS might be due to projection or magnetic fields (see Sect.~\ref{vrot_evol}). 
From Fig.~\ref{figure_9} we see that these stars (marked additionally by asterisks) 
also tend to show the lowest values of \eb\, for the corresponding mass range and
temperatures. Interestingly, the same is true for HD~93843 (marked by a diamond) 
which  is the only confirmed magnetic O-type star in our sample. Taken together, 
this might imply that extra broadening can be subject to projection effects, and 
that magnetic fields can play a role in determining the properties of this phenomenon. 
Evidence in support of the latter possibility has recently bee provided by \citet{jon13}.

\subsection{Interplay between rotation and extra broadening}
\label{rot_extra} 

\begin{figure}
{\includegraphics[width=8.9cm,height=6.5cm]{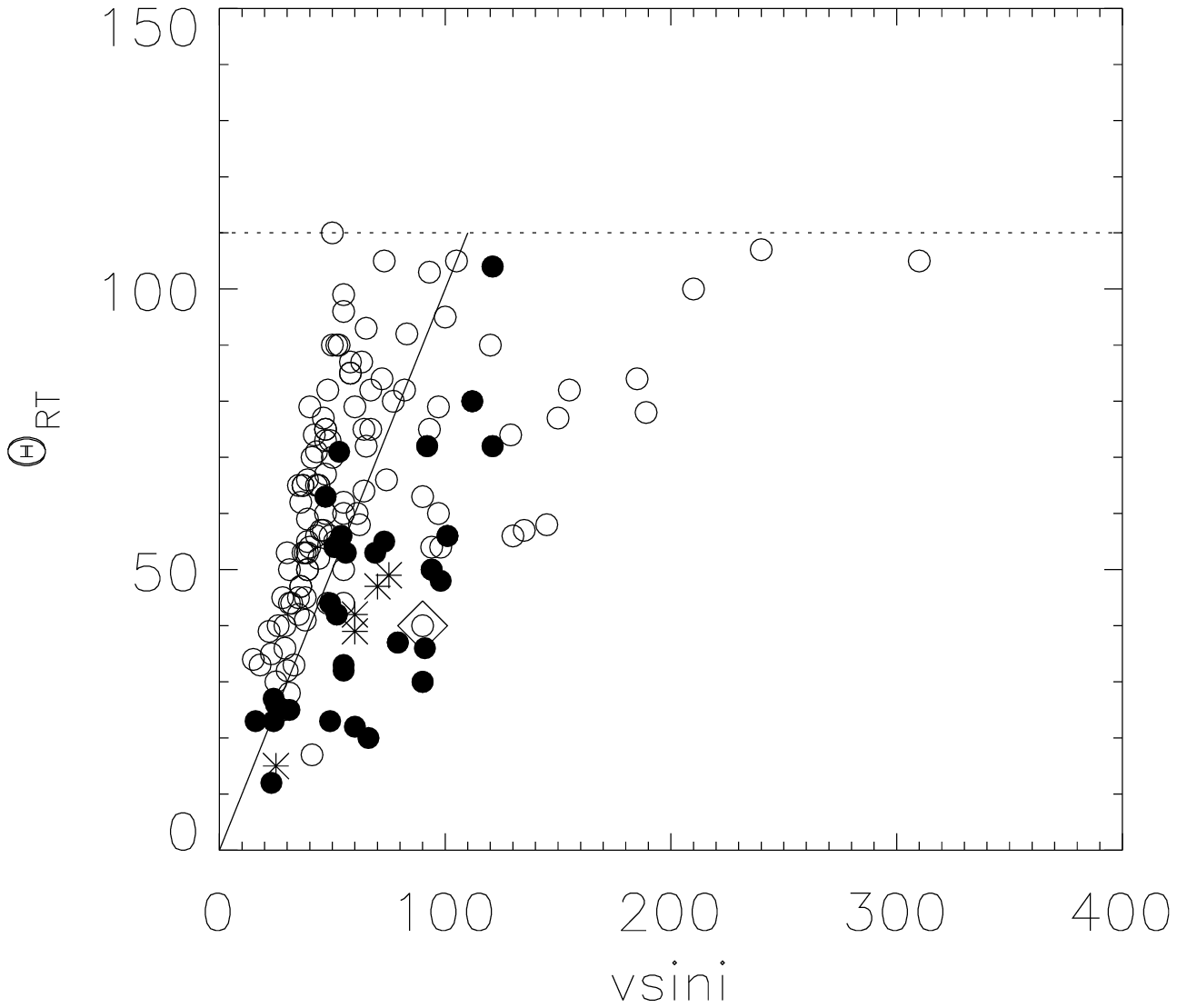}}
\caption{Extra line-broadening vs. projected rotational velocities for 
the stars from our OB sample (open circles) and the sample studied by 
Lefever et al. (2010, filled dots). Stars asumed to be affected by projection 
are highlighted by asterisks. The large diamond indicates the magnetic star 
HD~93843. The solid and dashed lines represent the one-to-one correspondence 
and the 110~\kms limit of \eb, respectively.}
\label{figure_10}
\end{figure}

From the analysis of the ESO O-star sample (Sect.~\ref{vmac_ostars}),
we found that \eb\, tends to increase with increasing \vsini. From the
larger dataset displayed in Fig.~\ref{figure_10}, we see that this
result is also valid for the complete OB sample (open circles and
asterisks), where the connection is now traced down to significantly
lower velocities. However, the relation is obviously not unique: stars
with the same or similar \vsini\, can show quite different values of
\eb, and vice versa. Although a direct influence of \vsini\, on \eb\,
cannot be excluded especially for very fast rotators, we find this to  
indicate that the two broadening mechanisms are tied together through 
similar dependences on stellar parameters and are not directly linked. 

For most of the sample stars with \vsini $\le$ 110~\kms, the extra
broadening either dominates or is in strong competition with rotation.
Interestingly, among the outliers one can note the objects that are 
assumed to be influenced by projection effects (asterisks) and the
magnetic star HD93843 (large diamond), which supports our previous
suggestion that extra broadening in hot massive stars is subject
to projection effects and effects caused by magnetic fields. 

In a previous study, \citet{lefever10} reported evidence for a positive 
correlation between \eb\, and \vsini\, for a large sample of low-mass 
B-stars (2~\Msun $\le$ \Minit $\le$ 6~\Msun) from the GAUDI database 
(Ground-based Asteroseismology Database Interface, \citealt{solano}). 
Since their estimates have been derived by applying a methodology similar 
to ours, we overplotted their data onto ours to allow for a direct comparison 
(filled dots in Fig.~\ref{figure_10}). Inspection reveals that while the 
\eb\, for the low-mass B-type stars indeed do increase with increasing 
\vsini, they are generally smaller than those observed for the massive 
OB stars at the same \vsini. Consequently, rotational broadening is stronger 
than the extra broadening for the majority of the low-mass stars.  Interestingly,
none of the targets studied by \citet{lefever10} shows extra broadening
in excess of 110~\kms.

Summarizing, we conclude that (i) extra broadening is most likely a 
fundamental feature of the spectra of OB stars with masses between 
$\sim$2 to $\sim$80~\Msun; (ii) rotational broadening is stronger than 
extra broadening in low-mass hot stars, whilst in high-mass ones it either 
dominates (for stars with \vsini $\ge$ 110~\kms)  or is of secondary importance 
(for stars with \vsini $\le$ 110~\kms) compared with the extra broadening;
(iii) objects with \eb\, in excess of 110~\kms are either very rare among 
OB stars or are not present at all.

\section{Summary}
\label{conclusions}

Based on our own data for a sample of 31 Galactiy O-type stars and 
incorporating similar data for 86 O stars and massive B-supergiants 
from the literature, we investigated the statistical properties of 
projected rotational and extra line-broadening rates as a function 
of stellar parameters and derived constraints on model predictions 
for the evolution of rotation in hot massive stars. 

The initial evolutionary masses of our sample stars range from 
$\sim$15 to  $\sim$80~\Msun, but the available data provide a good 
coverage from close to the ZAMS until the end of the B-supergiant phase 
only for \Minit $\gtrsim$ 40~\Msun,. Thus, the results summarized below 
can  be considered as representative for the more massive OB stars at
solar metallicity only. 

1. In good correspondence with results from \citet{simon11}), we found  
that  O-type stars, irrespective of their luminosity class, are 
subject to significant extra broadening. The impact of this broadening 
on the derived \vsini\, averaged over the  whole sample was estimated 
as -20$\pm$15~\kms, in perfect agreement with similar results from 
\citet{bouret12} regarding O-supergiants alone. 

2. Although our derived \vsini\, values were corrected for the effects 
of extra broadening, we repeated previous  findings (e.g. \citealt{penny96} 
and \citealt{howarth97}) that O stars and massive B supergiants show a 
clearly defined trend of increasing minimum \vsini\, with increasing \Minit, 
leading to a remarkable deficit of slowest projected rotators among the most 
massive objects. While differences in evolutionary stage can play a role to 
explain this puzzling problem, we suggested that either the axes of more 
massive OB supergiants are preferentially aligned with the Galactic poles, 
or that at the low-velocity regime there are additional (broadening) effects 
that influence the Fourier transform,  with microturbulence being a potential 
candidate. 

3. We found suggestive evidence that stars with \Minit $\gtrsim$ 50~\Msun\, 
rotate relatively slowly when appearing at or close to the ZAMS, with
velocities which do not exceed 26\% of their critical equatorial
velocities; for the majority, this limit might be even lower, about 17\%. 
Similar results have been reported by \citet{wolff06} for a sample of 
~55 Galactic O stars  with masses between 10 and 40~\Msun, and by 
\citet{ramirez} for a sample of 216 presumably single O-type stars in 30 
Dor in the Large Magellanic Cloud.  The possibility that massive stars 
can appear close to the ZAMS with velocities significantly lower than the 
critical speed may have important implications for the formation and evolution 
of these objects, including their potential to develop LGRB at the end
of their lives.

4. We confirmed evolutionary predictions from \citet{brott} that stars 
with \Minit $\ge$ 35~\Msun\, loose significant amounts of their angular
momentum when evolving from the ZAMS throughout the B-supergiant phase, 
with the BS jump playing an important role, especially for objects with 
high initial rotation. While for the stars on the hotter side of the BS 
jump the measured and the predicted rates also agree quantitatively, for 
those after the jump a significant discrepancy was detected, with the former 
being systematically higher than the latter. To reconcile theory with 
observations, we discussed two possibilities: either the observed 
\vsini\, are overestimated due to problems of the FT method (microturbulence?), 
and/or the angular momentum loss has been overpredicted, caused by overpredicted
mass-loss rates in the BS region. At least at present, we consider the second 
possibility as the more likely one.  

5. For massive B supergiants with \Teff $\le$ 4.25 dex, our analysis confirms  
prebious results about the lack of fast rotators. Following \citet{vink10} and 
using evolutionary tracks from \citet{brott}, we argued that for masses above  
$\sim$35~\Msun, BS braking can potentially  be responsible for the absence of 
fast-rotating stars at the cooler edge of the B-supergiant domain.

6. Described in terms of large-scale photospheric turbulence with a radial-
tangential distribution, the extra broadening velocities, \eb, of our sample stars 
were  found to be highly supersonic, with values ranging from $\sim$35 to 
$\sim$110~\kms. The main properties of this broadening, as derived in the present 
study, can be summarized as follows:

a) The \eb\, values of hot massive stars appear to decline toward cooler 
temperatures, later evolutionary stages, and lower initial mass. Marginal evidence 
that this broadening  is subject to projection and effects caused by magnetic fields 
was found. While the latter possibility is consistent with recent findings from 
\citet{jon13}, the former, if confirmed by future analyses, would hint at an aspheric 
origin of the extra broadening phenomenon. 

b) None of our sample stars shows \eb\, in excess of 110~\kms. Sincethe same limit 
applies to the stars with \Minit\, between 2 and 6~\Msun\, as studied by 
\citet{lefever10}, we suggested that this is a general feature of the extra 
broadening phenomenon in OB stars with $non-negligible$ extra broadening. The 
explanation of this result remains unclear so far.

c)  The extra broadening velocities of massive OB stars and of low-mass B-type 
stars show a tendency  to increase toward  higher \vsini. Because  stars with 
same or similar \vsini\,  can show quite different values of \eb, and vice versa, 
we suggested that this finding most likely reflects similar dependencies of \eb\, 
and \vsini\, on stellar parameters and is not indicative of a direct relation 
between them.  

d) Although on a general scale the extra broadening rates of more massive stars 
(\Minit $\ge$ 35~\Msun) decrease toward cooler temperatures,  there is suggestive 
evidence that this decline may depend on \Teff\, because it is negligible for 
log~\Teff~$\gtrsim$~4.3, but clearly notable below. Since the limiting temperatures 
of these two regimes coincide with the $observed$ position of the BS jump as 
determined by \citet{MP}, we suggested that extra broadening might be sensitive to  
changes of the wind properties when crossing the BS region.  

\acknowledgements{We thank the referee, Ian Howarth, for his valuable comments and 
suggestions. We are also thank Conny Aerts for stimulating discussions  about 
important issues concerning the analysis of extra broadening. NM acknowledges travel 
support from the German DAAD (grant number A/12/01733).}

\end{document}